\documentclass[journal]{IEEEtran}
\usepackage{times,amsmath,epsfig}
\usepackage{bbding}
\usepackage{epstopdf}
\usepackage{ulem}
\usepackage{caption}
\usepackage[ruled,vlined]{algorithm2e}
\usepackage{color}
\usepackage{algorithmic}
\usepackage{graphicx}
\usepackage{url}
\usepackage[bookmarks=false]{hyperref}
\usepackage{multirow}
\usepackage{multicol}
\usepackage{subfigure}
\usepackage{threeparttable}
\usepackage{booktabs}
\usepackage{ulem}
\usepackage{amsfonts}
\usepackage{textcomp}
\usepackage[framemethod=1]{mdframed}
\usepackage{enumitem}
\usepackage{amsthm}
\usepackage{scalerel}

\usepackage{amssymb} 
\usepackage{cite} 
\theoremstyle{plain}
\newtheorem{theorem}{Theorem}
\theoremstyle{definition}
\newtheorem{defi}[theorem]{Definition}
\theoremstyle{plain}
\newtheorem{lemma}[theorem]{Lemma}
\theoremstyle{plain}

\theoremstyle{plain}
\newtheorem{proposition}[theorem]{Proposition}

\ifodd 0
    \newcommand\rev[1]{{\color{blue}#1}}
    \newcommand{\com}[1]{\textbf{\color{red} (COMMENT: #1)}} 
\else
    \newcommand\rev[1]{{#1}}
    \newcommand{\com}[1]{}
\fi

\newcommand{\tabincell}[2]{\begin{tabular}{@{}#1@{}}#2\end{tabular}}

\title{\huge \textsf{DeepOPF}: A Feasibility-Optimized Deep Neural Network Approach for AC Optimal Power Flow Problems   
}

\author{Xiang~Pan, Minghua~Chen, Tianyu~Zhao, and Steven H. Low \thanks{Xiang Pan and Tianyu Zhao are with Department of Information Engineering, The Chinese University of Hong Kong. Minghua Chen is with School of Data Science, City University of Hong Kong. Steven H. Low is with Department of Computing and Mathematical Sciences and Department of Electrical Engineering, California Institute of Technology.  Corresponding author: Minghua Chen (minghua.chen@cityu.edu.hk).}
}

\begin{document}
\begin{NoHyper}
\maketitle

\begin{abstract}
To cope with increasing uncertainty from renewable generation and flexible load, grid operators need to solve alternative current optimal power flow (AC-OPF) problems more frequently for efficient and reliable operation. In this paper, we develop a Deep Neural Network (DNN) approach, called \textsf{DeepOPF}, for solving AC-OPF problems in a fraction of the time used by conventional iterative solvers. A key difficulty for applying machine learning techniques for solving AC-OPF problems lies in ensuring that the obtained solutions respect the equality and inequality physical and operational constraints. {Generalized a prediction-and-reconstruction procedure in our previous studies}, \textsf{DeepOPF} first trains a DNN model to predict a set of independent operating variables and then directly compute the remaining ones by solving the power flow equations. Such an approach not only preserves the power-flow balance equality constraints but also reduces the number of variables to be predicted by the DNN, cutting down the number of neurons and training data needed. {\textsf{DeepOPF} then employs a penalty approach with a zero-order gradient estimation technique in the training process towards guaranteeing the inequality constraints. We also drive a condition for tuning the DNN size according to the desired approximation accuracy, which measures its generalization capability.} It provides theoretical justification for using DNN to solve AC-OPF problems. Simulation results for IEEE 30/118/300-bus and a synthetic 2000-bus test cases demonstrate the effectiveness of the penalty approach. They also show that \textsf{DeepOPF} speeds up the computing time by up to two orders of magnitude as compared to a state-of-the-art iterative solver, at the expense of $<$0.2\% cost difference. 
\end{abstract}


\section*{Nomenclature}
{\small
\begin{IEEEdescription}[\IEEEusemathlabelsep\IEEEsetlabelwidth{MMM}]
    \item[Variable\ \ Definition]
    \item[$\mathcal{N}$] Set of buses, $N \triangleq \text{card}(\mathcal{N})$.
    \item[$\mathcal{G}$] Set of P-V buses. 
    \item[$\mathcal{D}$] Set of P-Q buses. 
    \item[$\mathcal{E}$] Set of branches.
    \item[$P_{Gi}$] Active power generation at bus $i$.
    \item[$P_{Gi}^{\min}$] Minimum active power generation at bus $i$. 
    \item[$P_{Gi}^{\max}$] Maximum active power generation at bus $i$.
    \item[$P_{Di}$] Active power load at bus $i$.
    \item[$Q_{Gi}$] Reactive power generation at bus $i$. 
    \item[$Q_{Gi}^{\min}$] Minimum reactive power generation at bus $i$. 
    \item[$Q_{Gi}^{\max}$] Maximum reactive power generation at bus $i$. 
    \item[$Q_{Di}$] Reactive power load at bus $i$. 
    \item[$V_i$] Complex voltage at bus $i$ (includes the magnitude $|V_i|$ and the phase angle $\theta_i$, i.e., $V_i=|V_i|\angle\theta_i$).
    \item[$V_i^{\min}$] Minimum voltage magnitude at bus $i$.
    \rev{\item[$V_i^{\max}$] Maximum voltage magnitude at bus $i$.}    
    \item[$y_{ij}$] Complex admittance of the branch $(i,j) \in \mathcal{E}$.
    \item[$S_{ij}^{\max}$] Transmission limit of the branch $(i,j) \in \mathcal{E}$.
\end{IEEEdescription}
We use $\text{card}(\cdot)$ to denote the size of a set. For P-Q buses, the corresponding generator output and the operating bound are set to $0$. Without loss of generality, we set bus $0$ as the slack bus.}
\section{Introduction}\label{sec:intro}
Optimal power flow (OPF) problem, first posed by Carpentier in 1962\cite{carpentier1962contribution}, is central to power system operation and concerns more than ten billion dollars each year in the U.S. alone~\cite{cain2012history}. It optimizes particular system objectives, e.g., generation cost, subject to power-flow balance and operational constraints regarding the generation, voltage, and branch flow. 

Conventionally, OPF problems are mainly concerned at the transmission grid level, and the distribution grids are {modeled} as passive load buses with aggregated demand from different users. The recent endeavors of grid modernization and accommodating high penetration of distributed energy resources (DERs), such as solar photovoltaic and wind generating units, convert the buses in the distribution grids into more active ones. While they provide enormous control flexibility to reduce the cost of grid operation, the intermittent and uncertain nature of DERs introduces significant fluctuations and randomness in power injections. To cope with the fluctuation and uncertainty, the grid operators usually solve stochastic OPF problems frequently to efficiently and reliably maintain the power supply and demand balance. Specifically, system operators often generate a large number of scenarios based on different load prediction (e.g., more than 1000) and solve the corresponding standard OPF problems (one for each scenario) in order to obtain a stochastically optimized solution~\cite{8309128}.

However, conventional iteration-based solvers may not be able to solve OPF problems fast enough for the purpose~\cite{zhang2020convex}. For example, suppose a single large-scale AC-OPF problem can be solved quickly (e.g., state-of-the-art solvers can return the solution in one second for power network with hundreds {of} buses~\cite{babaeinejadsarookolaee2019power}), solving 1000 AC-OPF problems needs more than 10 minutes, which can be too slow for adjusting system operating points in response to the changes of renewable power injection~\cite{tang2017real}. It is thus crucial to develop efficient OPF solvers for cost-effective and reliable modern power system operation.

Recently, learning-based approaches for solving OPF problems have received substantial attention. {The works in~\cite{deepopf1, deepopf2, deepopf+} develop the first DNN schemes to \textit{directly} generate feasible and close-to-optimal solutions for (security-constrained) DC-OPF problems in a fraction of the time used by iterative solvers. The idea is to leverage the universal approximation~\cite{hornik1991approximation} capability of DNNs to learn the mapping between load input and OPF solutions. Then one can pass the loads through the trained DNN to instantly obtain a quality solution. The works suggest the potential of using deep learning in solving OPF problems and attract a variety of further studies~\cite{9335481,owerko2020optimal,guha2019machine,zamzam2019learning,chatzos2020high,dobbe2019towards}, some beyond the OPF problem~\cite{donti2021dc3}.} Learning-based approaches have also been developed to facilitate the solving process for OPF problems, by, e.g., determining active/inactive constraints to reduce the problem size\cite{zhang2020convex,deka2019learning, ng2018statistical,chen2020learning,9034123,robson2019learning} or speeding up the iterations in conventional solvers~\cite{biagioni2019learning,baker2019learning,diehl2019warm,dong2020smart,baker2020learning,zhang2021learning}. We present detailed discussions of related works in Sec.~\ref{related.work}. 

\begin{table*}[!t]
	\centering
	\caption{Summary of existing studies on machine learning for solving OPF problems.} 
    \renewcommand{\arraystretch}{1.1}
 		\begin{threeparttable}[b]
			\begin{tabular}{c|c|c|c|c|c|c|c}
		    \toprule			
				\hline
				\multirow{2}{*}{\tabincell{c}{Category}} &				
				\multirow{2}{*}{\tabincell{c}{Approach}} &
				\multirow{2}{*}{\tabincell{c}{Existing Study}} &
				\multicolumn{2}{c|}{\tabincell{c}{Problem}} &
				\multicolumn{3}{c}{\tabincell{c}{Metrics in Consideration}} \\
				\cline{4-8}
                &&&\tabincell{c}{DC-OPF} & \tabincell{c}{AC-OPF}
				&\tabincell{c}{Feasibility} & \tabincell{c}{Optimality} &\tabincell{c}{Speedup} \\
				\hline
                \multirow{6}{*}{\tabincell{c}{Hybrid}}&\multirow{2}{*}{\tabincell{c}{\tabincell{c}{Determining active constraints}}}&~\cite{deka2019learning}&\checkmark&&\XSolidBrush&\checkmark&\checkmark\\
                \cline{3-8}               
                &&~\cite{ng2018statistical,chen2020learning,zhang2020convex}&\checkmark&&\checkmark&\checkmark&\checkmark\\
                \cline{2-8}
                &\multirow{2}{*}{\tabincell{c}{Determining inactive constraints}}&~\cite{9034123}&\checkmark&&\XSolidBrush&\checkmark&\checkmark\\
                \cline{3-8}
                &&~\cite{robson2019learning}&&\checkmark&\XSolidBrush&\checkmark&\checkmark\\
                \cline{2-8}                
                &\multirow{2}{*}{\tabincell{c}{Predicting warm-start point}}&~\cite{biagioni2019learning} &\checkmark&&\checkmark&\checkmark&\checkmark\\
                \cline{3-8}                
                &&~\cite{baker2019learning,diehl2019warm,dong2020smart,zhang2021learning} &&\checkmark&\checkmark&\checkmark&\checkmark\\
                \cline{2-8}                
                &\multirow{1}{*}{\tabincell{c}{Predicting gradient in iterative algorithms }}&~\cite{baker2020learning} &&\checkmark&\XSolidBrush&\checkmark&\checkmark\\
                \hline
                \multirow{3}{*}{\tabincell{c}{Stand-alone}}&\multirow{3}{*}{\tabincell{c}{\rev{Learning the load-solution mapping and} \\ \rev{generating solutions directly from load inputs}}}&~\cite{deepopf1,deepopf2,deepopf+,zhao2021ensuring,9335481} &\checkmark&&\checkmark&\checkmark& \checkmark\\
                \cline{3-8}                
                &&~\cite{owerko2020optimal,guha2019machine,zamzam2019learning,chatzos2020high,dobbe2019towards,9115822,singh2021learning,fioretto2020predicting}&&\checkmark&\XSolidBrush&\checkmark&\checkmark\\
                \cline{3-8} 
				 &&\textbf{This work} &&\checkmark&\checkmark&\checkmark&\checkmark\\
				\hline
			\end{tabular}
 	\end{threeparttable}
	\label{table1}
\end{table*}

In this paper, we generalize the 2-stage approach in~\cite{deepopf1, deepopf2} to develop a DNN scheme for solving AC-OPF problems directly.\footnote{We note that there are also two independent works~\cite{guha2019machine} and~\cite{zamzam2019learning} applying the 2-stage approach in~\cite{deepopf1} to solve AC-OPF problems. We discuss the difference and similarity in Sec.~\ref{related.work} and compare the performance in Sec.~\ref{sec:simulations}.} As compared to using DNN to solve DC-OPF problems~\cite{deepopf1, deepopf2}, developing DNN schemes for AC-OPF problems face the following unique challenges.
\begin{itemize}
\item It is non-trivial to ensure the \textit{feasibility} of the generated solution, i.e., ensuring the DNN solution satisfies the non-convex power-flow balance equations and the operation limits simultaneously.

\item It is challenging to guide the DNN design (i.e., setting the number of hidden layers and the number of neurons per layer) to achieve the desired performance.
\end{itemize}
In this paper, we carry out a comprehensive study on the above challenges and make the following contributions.

$\rhd$ After briefly reviewing AC-OPF problem in Sec.~\ref{sec: OPF.review}, we develop a DNN approach for solving AC-OPF problem directly by generalizing the 2-stage Predict-and-Reconstruct (PR2) framework in~\cite{deepopf1, deepopf2} to AC-OPF setting in Sec.~\ref{ssec:Prediction and Reconstruction}. To guarantee the power-flow balances, we first train a DNN model to predict a set of independent operating variables and reconstruct the remaining ones by solving the AC power flow equations. We then employ a penalty approach in training the DNN so that the reconstructed solutions (generation, voltages, and branch flows) respect the corresponding operation limits. Due to the non-linearity of the AC power flow equations, it is difficult to compute the penalty gradient. We further apply a zero-order optimization technique in the training algorithm to compute the penalty gradients efficiently.

$\rhd$ We characterize the load-to-solution mapping and study how well could a DNN learn such a mapping in Sec.~\ref{sec:OPF.theory_analysis}. We first show that the load-to-solution mapping is continuous and differentiable almost everywhere when the optimal solution is unique, adding a new understanding to the AC-OPF literature. Consequently, by the universal approximation capability~\cite{hornik1991approximation}, DNN can approximate the load-to-solution mapping arbitrarily well as the number of neurons increases. We further derive a condition for tuning the size of the DNN according to the desired approximation accuracy. To our best knowledge, these results provide the first theoretical justification of using DNN to solve AC-OPF problem.

$\rhd$ We carry out simulations on IEEE 30/118/300-bus and a synthetic 2000-bus test cases and summarize the results in Sec.~\ref{sec:simulations}. \textsf{DeepOPF} speeds up the computing time by up to two orders of magnitude with $<$0.2\% cost difference as compared to Pypower~\cite{tpcwTrey1}. {The comparison results with recent learning-based schemes show the effectiveness of the penalty approach in improving DNN solutions' feasibility.} We also observe that the DNN model trained with the estimated gradient from zero-order techniques achieves a similar or better feasibility rate than that trained with exact gradient by exploiting the implicit function theorem (see, e.g.~\cite{donti2021dc3}), which can be of independent interest. \rev{Due to the space limitation, all proofs are in the technical report~\cite{deepopf3}.}
 
\section{Related Work}\label{related.work}
{
Leveraging learning techniques to solve the OPF problem is becoming an active area of research. 
As seen in Table~\ref{table1}, existing supervised-learning based works can be grouped into two \textit{orthogonal} categories: the hybrid approach~\cite{zhang2020convex,deka2019learning,ng2018statistical,chen2020learning,9034123,robson2019learning,biagioni2019learning,baker2019learning,diehl2019warm,dong2020smart,baker2020learning,zhang2021learning} and the stand-alone approach~\cite{deepopf1,deepopf2,deepopf+,9335481,owerko2020optimal,guha2019machine,zamzam2019learning,chatzos2020high,dobbe2019towards,9115822,singh2021learning,fioretto2020predicting}. The hybrid approach applies the learning technique to accelerate the solving process of conventional methods. Some works for DC-OPF focus on learning the active/inactive constraints that can maintain the optimality and achieve speedup by reducing the problem size~\cite{zhang2020convex,deka2019learning,ng2018statistical,chen2020learning,9034123,robson2019learning}. Others predict warm-start points or gradient for iteration to accelerate the solving process by providing initial points close to the optimal or replacing the computationally expensive gradient computation with a fast neural-network predictor~\cite{biagioni2019learning,baker2019learning,diehl2019warm,dong2020smart,baker2020learning,zhang2021learning}. The hybrid approach's limitations lie in the limited speedup performance due to the inevitable iteration process. For example, the speedup performance reported in~\cite{baker2019learning,diehl2019warm,dong2020smart} for the hybrid approach predicting warm-start points is up to $4\times$ for a 2000-bus test case while the stand-alone scheme reported in this paper achieves up to $123\times$ speedup for the same 2000-bus test case.

The stand-alone approach directly generates solutions to the OPF problem. 
Some works~\cite{owerko2020optimal,chatzos2020high,dobbe2019towards,9115822,fioretto2020predicting,singh2021learning} use NN to predict all the variables and model the constraint violation into the loss function. Yet the obtained solutions suffer from the in-feasibility issue as the equality constraints may not be satisfied due to prediction errors. Meanwhile,~\cite{deepopf1,deepopf2,deepopf+} propose a PR2 framework for solving the (security-constrained) DC-OPF problem, which predicts a subset of the variables and reconstructs the remaining ones by leveraging the equality constraints. They also integrates the penalty term into the training process to pursue the inequality constraints satisfaction. The authors in~\cite{zamzam2019learning} apply the PR2 framework to design a DNN to generate solutions to AC-OPF problems, but it does not consider the line limit constraints 
and the generated solutions may not be feasible. ~\cite{fioretto2020predicting} also designs a framework similar to the PR2 one to solve AC-OPF problem, but it does not consider the inequality constraints' violation and may result in in-feasible solutions.

Besides the supervised-learning approach, there is an emerging line of research~\cite{donti2021dc3, deepopfngt,zhou2020deriving} on developing an unsupervised/reinforcement learning framework for solving the OPF problem. For example,~\cite{donti2021dc3} trains a DNN to minimize the generation cost and the penalty regarding constraints violations, which exploit the implicit function theorem to compute the penalty gradient for tuning the DNN's parameters. Although both works do not need to prepare the ground truth, which can be computational expensive, they have some certain limitations.~\cite{donti2021dc3} does not consider the line limit constraints, and~\cite{deepopfngt, deepopfv} may lead to load dissatisfaction. 

In this paper, we generalize the PR2 framework in our previous work~\cite{deepopf1,deepopf2,deepopf+} and integrate the penalty approach for solving AC-OPF problems directly. The penalty approach has been used in~\cite{deepopf1,deepopf2,deepopf+,chatzos2020high,9115822,donti2021dc3,fioretto2020predicting}. A key challenge of applying the penalty approach to the AC-OPF setting lies in that there does not exist an explicit expression between the predicted variables and the reconstructed variables due to the non-linear AC power flow equations for gradient-based training method. This makes it difficult to compute the penalty gradient. One contribution of this paper is to address the challenge of obtaining the penalty gradients efficiently by a zero-order technique during training, to improve the feasibility performance. We also evaluate the performance of \textsf{DeepOPF} on medium-/large-scale cases with up to 2000 buses.}

\section{AC-OPF Problem} \label{sec: OPF.review}
We study the standard AC-OPF problem with the bus injection model\footnote{There are two equivalent models in the OPF problem, namely the bus injection model and the branch flow model~\cite{subhonmesh2012equivalence}. The \textsf{DeepOPF} approach applies to both models. We focus on the bus injection model in this paper.} as follows: 
\begin{eqnarray}
    &{\min} &\sum\nolimits_{i=1}^{N}{C_i\left( P_{Gi} \right)} \label{obj}\\
    &\mathrm{s.t.}& \label{acopf_con1}
    \sum_{\left( i,j \right) \in \mathcal{E}}{\mathrm{Re}\left\{ V_i\left( V_{i}^{*}-V_{j}^{*} \right) y_{ij}^{*} \right\} =P_{Gi}-P_{Di}}, \;\;\;\\
    \label{acopf_con2}
    &\ &\sum_{\left( i,j \right) \in \mathcal{E}}{\mathrm{Im}\left\{ V_i\left( V_{i}^{*}-V_{j}^{*} \right) y_{ij}^{*} \right\} =Q_{Gi}-Q_{Di}}, \;\;\;\\
    \label{acopf_con3}
    &\ &P_{Gi}^{\min}\le P_{Gi}\le P_{Gi}^{\max}, i\in \mathcal{N}, \\
    \label{acopf_con4}
    &\ &Q_{Gi}^{\min}\le Q_{Gi}\le Q_{Gi}^{\max}, i\in \mathcal{N},\\
    \label{acopf_con5}
    &\ &V_{i}^{\min}\le |V_i| \le V_{i}^{\max}, i\in \mathcal{N},\\
    \label{acopf_con6}
    &\ &| V_i\left( V_{i}^{*}-V_{j}^{*} \right) y_{ij}^{*} |\le S_{ij}^{\max},\left( i,j \right) \in \mathcal{E},\\
    \label{acopf_con7}
    &\mathrm{var.}& P_{Gi},Q_{Gi},V_i,i\in \mathcal{N}, \nonumber
\end{eqnarray}
\noindent where $\text{Re} \{z\}$, $\text{Im}\{z\}$, and $z^*$ denote the real part, imaginary part, and the conjugate of $z$, respectively. The typical objective is to minimize the total cost of active power generations, where $C_i(\cdot)$ is the individual generation cost function and is commonly quadratic. The constraints are the power-flow balance equations in \eqref{acopf_con1} and \eqref{acopf_con2}, the active and reactive generation limits in \eqref{acopf_con3} and \eqref{acopf_con4}, the voltage magnitude limit in \eqref{acopf_con5}, and the branch flows limits in \eqref{acopf_con6}. {We note that $(P_{D_i}, Q_{D_i})$ is the net load of bus $i$, which is the difference between the actual electricity demand and the DERs power generation, e.g., solar or wind generation, at the bus.

As discussed in Sec~\ref{sec:intro}, grid operators may need to solve stochastic AC-OPF problems in practice, which requires one to solve a large number of standard AC-OPF problems efficiently. However, the AC-OPF problem is NP-hard~\cite{bienstock2019strong} and non-convex. Consequently, no solvers can solve general AC-OPF problems exactly in polynomial time unless P = NP.\footnote{Note that AC-OPF problems under specific settings may still be polynomial-time solvable. For example, some existing works, e.g.,~\cite{low2014convex2}, characterize sufficient conditions under which AC-OPF problems can be solved exactly in polynomial time by convexification.} This observation motivates the studies of reducing the time for solving AC-OPF problems. 

Recently, there has been active research in employing DNN to solve OPF problems directly, in a fraction of the time used by iterative solvers; see detailed discussions in Sec.~\ref{related.work}. The idea is to leverage the approximation capability of DNNs ~\cite{hornik1991approximation} to learn the load-solution mapping of the OPF problem. Then one can feed the load to the DNN to obtain a solution instantly. In this paper, we generalize the 2-stage approach in~\cite{deepopf1, deepopf2} to develop a DNN scheme to learn the load-solution mapping for solving AC-OPF problems directly with optimized solution feasibility.\footnote{We note that the non-convex AC-OPF problem may admit multiple optimal solutions, thus multiple load-solution mappings. Meanwhile, it is empirically observed that (i) there is a unique solution that satisfies the operational constraints given a ``reasonable'' load region~\cite{tang2017real} and (ii) current state-of-the-art solvers can obtain close-to-optimal OPF solutions. Based on these observations, we assume the load-solution mapping generated by the state-of-the-art solver (Pypower in our case) is unique and regarded as the mapping to be learned by the DNN scheme. 
}} 


\section{A Feasibility-Optimized Deep Neural Network Approach For AC-OPF} \label{sec: PR2_DNN}
\subsection{Overview of \textsf{DeepOPF}}\label{ssc:overview_of_deepopf}
Fig.~\ref{fig1} presents the predict-and-reconstruct (PR2) framework of \textsf{DeepOPF}. The idea is to train a DNN model to predict a set of independent operating variables and then directly compute the remaining dependent variables by solving AC power-flow equations. The proposed \textsf{DeepOPF} guarantees that the power-flow equality constraints are satisfied and reduces the dimension of mapping to learn, subsequently cutting down the size of the DNN and the amount of training data needed.

While the power flows are balanced by following the PR2 framework, the fundamental difficulty lies in ensuring that the obtained solutions satisfy generations' operation limits, voltages, and branch flows constraints. Most recent studies in this area for AC-OPF did not consider this hurdle for ensuring feasibility, as discussed in Table~\ref{table1}. In general, there exist two methods to tackle this issue. One is to extend the preventive learning framework for DC-OPF~\cite{deepopf+}, where we strengthen the operating constraints used in training, therefore anticipating the resulting predicted solutions remain feasible to the DC-OPF problem with default constraints even with approximation errors. However, it is non-trivial to determine how much we should calibrate the operating constraints for AC-OPF problem without reducing the load input region. The other is to integrate our PR2 framework with the penalty approach~\cite{deepopf2} by adding the constraints-related term in the loss function during the training process. The penalty approach excels in that it does not need to reduce the supportable load region. However, as we will see in Sec.~\ref{ssec:zoo.alg}, it requires us to run a gradient-like algorithm without access to the gradient information directly. The proposed \textsf{DeepOPF} leverages zero-order optimization techniques to address the issue.

\begin{figure}[!t]
	\centering
	\includegraphics[width = 0.48\textwidth]{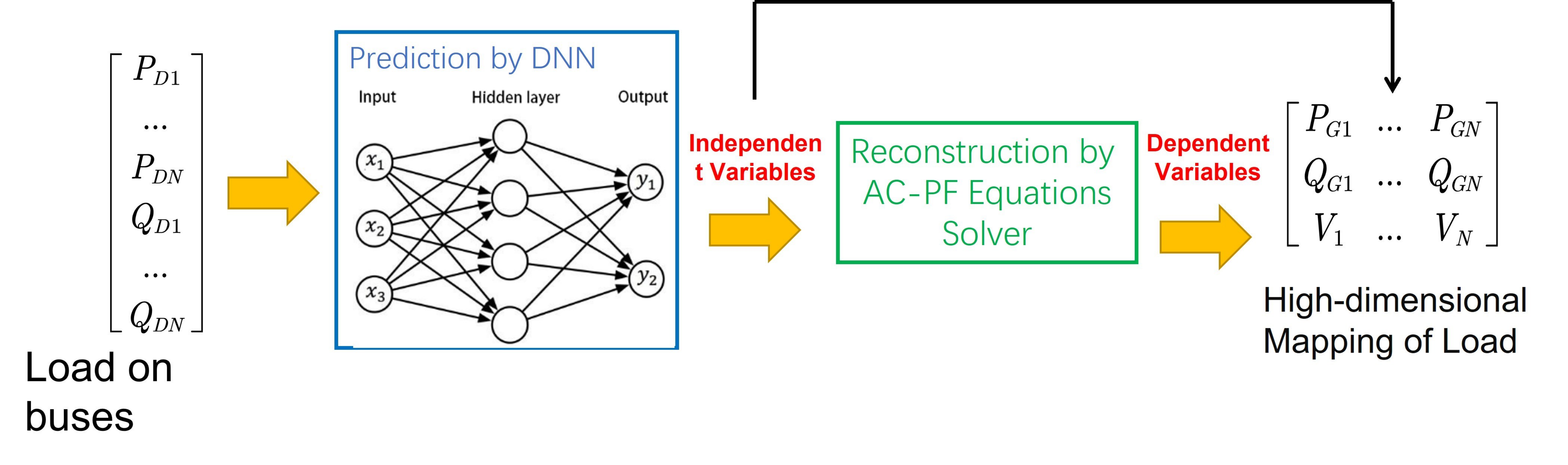}
	\captionsetup{font={small}}
	\caption{The predict-and-reconstruct framework for designing DNN solvers for AC-OPF problems. \protect\footnotemark The DNN is trained to predict a selected set of independent variables, as listed in Table~\ref{table2}. The remaining dependent variables are reconstructed via solving the non-linear AC power-flow equations.}
	\label{fig1}
\end{figure}
\footnotetext{The framework is generalized from the one for DC-OPF problems proposed in~\cite{deepopf1,deepopf2}, by replacing the DC-PF equations with AC power flow equations. \rev{It is also used in two independent works~\cite{zamzam2019learning} and~\cite{guha2019machine}.}}

\vspace{-0.1in}
\subsection{Data Preparation}\label{ssec:load.sampling.and.pre-preprocessing}
To train and test the DNN model, we sample the training and test load data, i.e., $P_{D_i}$ and $Q_{D_i},i\in \mathcal{N}$ within a given range of the default value, uniformly at random, which helps avoid the over-fitting issue. The sampling data is then fed into the traditional AC-OPF solver to obtain reference solutions. Following the common practice, we normalize each dimension of training data with the standard variance and mean of the corresponding dimension before training.

\vspace{-0.1in}
\subsection{Prediction and Reconstruction} \label{ssec:Prediction and Reconstruction}
We summarize the set of independent variables (to predict) and the remaining dependent variables (to reconstruct) for each type of bus in Table~\ref{table2}. 
As seen, DNN predicts the voltage phase angle and voltage magnitude on the slack bus, $\theta_0$ and $|V_0|$, and the set of the active power generation and the voltage magnitude for the P-V buses, i.e., $P_{Gi}$ and $|V_i|,i\in \mathcal{G}$. Every variable in Table~\ref{table2} to be predicted by the DNN, denoted by $x_{pred}$, is related to a operating constraints, i.e., $x^{\min} \le x_{pred} \le x^{\max}$. Similar to \cite{deepopf1,deepopf2,zamzam2019learning}, we associate with it a one-to-one corresponding scaling factor by
\begin{equation}
    x_{pred}=s_{pred} \cdot \left( x^{\max}-x^{\min} \right)+ x^{\min},
\label{scaling}
\end{equation}
\noindent where $s_{pred}$ is the scaling factor. The DNN will predict $s_{pred} $ and compute $x_{pred}$ by \eqref{scaling}. In our design, the Sigmoid function~\cite{goodfellow2016deepma} is applied as the activation function of the output layer to ensure the predicted scaling factor to within $(0, 1)$.

We reconstruct the remaining dependent variables listed in Table~\ref{table2} by solving the nonlinear AC power-flow equations by the widely-used Newton's method~\cite{mehta2016recent}, with the predicted independent variables as inputs. It is well understood that there exist multiple solutions to the AC power flow equations. Newton's method or similar iterative algorithms generate different solutions to AC power flow equations using different initial points~\cite{mehta2016recent}. \rev{Meanwhile, the convergence of Newton's method also depends on the chosen initial points. In our design, we set Newton's method's initial points as the average values of the dependent variables in the historical/training data. In our simulations in Sec.~\ref{sec:simulations}, we observed that Newton's method with such initialization always converges.}

\vspace{-0.1in}
\subsection{DNN Model} \label{ssec:DNN Model}
In \textsf{DeepOPF}, we design the DNN model based on the multi-layer feed-forward neural network structure:
\begin{eqnarray}
    \label{dnn1}
    &\mathbf{s}_0&=[\mathbf{P}_D,\mathbf{Q}_D],\\
    \label{dnn2}
    &\mathbf{s}_i&=\sigma \left( W_i\mathbf{s}_{i-1}+\mathbf{b}_{i} \right),\forall \ i=1, ..., L,\\
    \label{dnn3}
    & \mathbf{s}_{pred} &=\sigma '\left( W_{L+1} \rev{\mathbf{s}_L}+\mathbf{b}_{L+1} \right),
\end{eqnarray}
where $\mathbf{P}_D =\left[ P_{D_i},i\in \mathcal{N} \right]$ and $\mathbf{Q}_D=\left[ Q_{D_i},i\in \mathcal{N} \right]$ are the active and reactive load vector, consisting DNN's input $\mathbf{s}_0$. $\mathbf{s}_i$ is the output vector of the $i$-th hidden layer, depending on matrices $W_i$, biases vectors $\mathbf{b}_i$ and the ($i-1$)-th layer's output $\mathbf{s}_{i-1}$. $W_i$ and $\mathbf{b}_i$ are subject to the DNN design. $L$ is the number of hidden layer. $\sigma (\cdot)$ and $\sigma '(\cdot)$ are ReLU activation function used in the hidden layers and the Sigmoid action function used in output layer, respectively. In our simulation, we use a DNN with \rev{two} hidden layers; thus $L = 2$. The DNN predicts voltage magnitude on the slack bus, the active power generation, and the voltage magnitude on the P-V buses. Thus, the output dimension of DNN model is ${2\cdot\text{card}(\mathcal{G})+1}$.

\begin{table}[!t]
	\renewcommand{\arraystretch}{1.1}
	\caption{The selected independent and dependent variables.}
	\centering
	\begin{tabular}{c|c|c|c}
		\hline
		{Type of bus} & {Slack} & {P-Q} & {P-V} \\
		\hline
		\tabincell{c}{Set of independent \\variables} & \tabincell{c}{$\theta_0$,$|V_0|$} & \tabincell{c}{$P_{Di}$,$Q_{Di}$,\\$i\in \mathcal{D}$} & \tabincell{c}{$P_{Gi}$,$|V_i|$,\\$i\in \mathcal{G}$}\\
		\hline
		\tabincell{c}{Set of dependent \\variables} & \tabincell{c}{$P_{G0},Q_{G0}$} & \tabincell{c}{$\theta_i, |V_i|$,\\$i\in \mathcal{D}$} & \tabincell{c}{$\theta_i, Q_{Gi}$,\\$i\in \mathcal{G}$} \\
		\hline
	\end{tabular}
	\label{table2}
\end{table}

\subsection{Penalty Approach based Training Scheme}\label{ssec:penalty.approach.traiing}
After constructing the DNN model, we design a loss function to guide the training. For each instance in the training data set, the loss function consists of two parts. The first part is the prediction error, which is computed with the squared $\ell_2$ norm between the generated scaling factors vector $\mathbf{s}_{pred}$ and the reference vector $\mathbf{s}_{ref}$:
\begin{eqnarray}
\mathcal{L}_{pred}=\frac{1}{{2\cdot \text{card}(\mathcal{G})+1}}\lVert \mathbf{s}_{pred}-\mathbf{s}_{ref} \rVert_{2}^{2}.
\label{loss_prediction_error}
\end{eqnarray}
The second part is a penalty term to capture the violation of the inequality constraints in AC-OPF formulation, including the generation limits, the voltage magnitude constraints, and the branch flow limits.\footnote{We note that by using the Predict-and-Reconstruct framework, \textsf{DeepOPF} guarantees the power-flow balance equality constraints always satisfied.} For the reconstructed variable in Table~\ref{table2}, say $x_{rec}$, the corresponding penalty is defined by:
\begin{equation}
    p\left( x_{rec} \right) =\max \left( x_{rec}-x_{rec}^{\max},0 \right) +\max \left( x_{rec}^{\min}-x_{rec},0 \right),
    \label{penalty}
\end{equation}
which gives a positive penalty if $x_{rec}$ is outside the feasible region $\left[x_{rec}^{\min}, x_{rec}^{\max}\right]$ and zero otherwise. The overall penalty in the loss function is the average penalty of the reconstructed variables as follows:
\begin{align}
    \mathcal{L}_{pen}=&\frac{1}{\text{card} (\mathcal{E})}\sum_{\left( i,j \right) \in \mathcal{E}}{p\left( S_{rec,ij} \right)}+\frac{1}{\text{card} (\mathcal{D})}\sum_{i\in \mathcal{D}}{p\left( |V_{rec,i}| \right)} \nonumber \\
    +&\frac{1}{\text{card}(\mathcal{G})}\sum_{i\in \mathcal{G}}{p\left( Q_{rec,Gi} \right)}+p\left( P_{rec,G0} \right)+p\left( Q_{rec,G0} \right),
    \label{loss_penalty}
\end{align}
\noindent where $p\left( S_{rec,ij} \right)$, $p\left( |V_{rec,i}| \right)$, $p\left( Q_{rec,Gi} \right)$, $p\left(P_{rec,G0} \right)$ and $p\left( Q_{rec,G0} \right)$ represent the violation for each reconstructed branch flow, the violation for each reconstructed voltage magnitude at P-Q buses, the violation for each reconstructed reactive power generation at P-V buses, and the violation for the reconstructed active/inactive power generation at the slack bus, respectively. The overall loss function is given as:
\begin{eqnarray}
    \mathcal{L}=w_1\cdot \mathcal{L}_{pred}+w_2\cdot \mathcal{L}_{pen},
\label{total.loss}
\end{eqnarray}
\noindent where $w_1$ and $w_2$ are positive weighting factors used to balance the prediction error and the penalty during the training. The training processing is to minimize the average loss of the training data by adjusting the DNN's parameters $W_i$ and $\mathbf{b}_i$:
\begin{equation}
    \min_{W_i,\mathbf{b}_i,i=1,...,L}\frac{1}{\mathrm{card}\left( \mathcal{T} \right)}\sum_{k\in \mathcal{T}}{\mathcal{L}_{k}},
    \label{loss_func}
\end{equation}
\noindent where $\mathcal{T}$ is the training data-set, and $\mathcal{L}_{k}$ is the loss of the training data with index $k$. 

\vspace{-0.1in}
\subsection{Zero-order Optimization Technique for Penalty Approach}\label{ssec:zoo.alg}
It is common to apply the first-order gradient-based schemes, e.g., the stochastic gradient descent (SGD) algorithm, to solve~\eqref{loss_func} in training. This requires the gradients information i.e., $\nabla \mathcal{L}_{pred}$ and $\nabla\mathcal{L}_{pen}$, with respect to $W_i$ and $\mathbf{b}_i$.
We can first compute $\nabla \mathcal{L}_{pred}$ and $\nabla\mathcal{L}_{pen}$ w.r.t DNN's output ($\mathbf{s}_{pred}$), and apply chain rule to get the gradients w.r.t $W_i$ and $\mathbf{b}_i$. As $\mathcal{L}_{pred}$ only depends on $\mathbf{s}_{pred}$, we directly compute the corresponding $\nabla \mathcal{L}_{pred}\left( \mathbf{s}_{pred} \right)$ according to \eqref{loss_prediction_error}. However, it is hard to compute $\nabla\mathcal{L}_{pen}\left( \mathbf{s}_{pred} \right)$ as there does not exist an explicit expression between the predicted variables and the reconstructed variables, making it difficult to compute the penalty gradient w.r.t. the DNN's output directly. It is the key challenge in applying the penalty approach in training DNN for solving AC-OPF problems. 

We address the challenge as follows. We estimate the gradients $\nabla\mathcal{L}_{pen}$ w.r.t DNN's output by a two-point zero-order optimization technique~\cite{AgarwalDX10,9186148} as: 
\begin{align}
    \hat{\nabla}\mathcal{L}_{pen}\left( \mathbf{s}_{pred} \right) &= \nonumber \\
    \frac{d \cdot \mathbf{v}}{2\delta}[& \mathcal{L}_{pen}\left( \mathbf{s}_{pred}+\mathbf{v}\delta \right) -\mathcal{L}_{pen}\left( \mathbf{s}_{pred}-\mathbf{v}\delta \right) ], \label{zero_order_estimation}
\end{align}
\noindent where $d$ is the output dimension of the DNN model, $\delta>0$ is a smooth parameter, and $\mathbf{v}$ is a vector sampled uniformly at random on the unit ball. The following proposition shows the effectiveness of the zero-order method for gradient estimation.
\begin{proposition} \label{prop1}
    Assumed $\mathcal{L}_{pen}$: $\mathbb{R}^d\mapsto \mathbb{R}$ is differentiable, the gradient estimator in~\eqref{zero_order_estimation} is unbiased, i.e., 
    \begin{equation*}
        \lim_{\delta\rightarrow 0}\mathbf{E}_{\mathbf{u}}\left( \hat{\nabla}\mathcal{L}_{pen}\left( \mathbf{s}_{pred} \right) \right) =\nabla \mathcal{L}_{pen}\left( \mathbf{s}_{pred} \right). 
    \end{equation*}        
\end{proposition}
\vspace{-0.1in}
\noindent We discuss the advantage of the two-point zero-order optimization technique in the following. First, it is more efficient than the standard gradient estimation method. The method would solve the AC power flow equations many times by uniform sampling on the unit ball so as to compute gradient numerically. It is computationally expensive. In contrast, the two-point gradient estimator only requires solving the AC power flow equations twice regardless of the input space dimensions, which substantially reduces the computation burden. Second, the two-point scheme is more efficient than other zero-order optimization techniques, e.g., one-point or multi-point methods~\cite{9186148}, in that the two-point scheme usually achieves a better trade-off between computation efficiency and estimation variance in practice. Further, it is straightforward to implement the gradient estimator in~\eqref{zero_order_estimation}.

\rev{After obtaining $\hat{\nabla}\mathcal{L}_{pen}\left( \mathbf{s}_{pred} \right)$, we compute the gradient for the entire loss w.r.t. the output of DNN as $\hat{\nabla}\mathcal{L}\left( \mathbf{s}_{pred} \right) =w_1\nabla \mathcal{L}_{pred}\left( \mathbf{s}_{pred} \right) +w_2\hat{\nabla}\mathcal{L}_{pen}\left( \mathbf{s}_{pred} \right) $. We then use the back-propagation algorithm~\cite{goodfellow2016deepma} to compute the loss gradient with respect to $W_i$ and $\mathbf{b}_i$ for the $i$th layer in the DNN model. The outline of the process is summarized in Algorithm~\ref{alg:zoo.adam}. 

We note that the proposed algorithm is essentially a variant of SGD algorithms. Research~\cite{ghadimi2013stochastic} suggests the following general conditions for the convergence of SGD algorithms: (i) The gradient estimation is an unbiased approximation of the exact gradient; (ii) The variance of estimation error is upper-bounded by a constant; (iii) The exact gradient is Lipchitz-continuous w.r.t the input. In our setting, condition (i) holds according to Proposition~\ref{prop1}. However, it is difficult to check whether conditions (ii) and (iii) hold without the explicit expression of the penalty gradient, which we leave as a future direction.\footnote{\rev{Although one can still compute the penalty gradient by applying the implicit function theorem (see, e.g., ~\cite{donti2021dc3}), in the absence of an explicit expression of the penalty function to the predicted variables, the computation involved in checking the two conditions are non-trivial. In particular, the computation involves the inversion operation for the Jacobian matrix of the reconstructed variables w.r.t. the predicted variables, which is not a constant matrix. To check conditions (ii) and (iii), we need to obtain the upper bound of the norm of the exact gradient. This in turn needs us to bound the maximum singular value of the inverse Jacobian matrix, which is non-trivial as we need to solve a non-convex optimization to obtain the bound. 
}} Instead, we develop a DNN training algorithm using the two-point gradient estimation technique and empirically observe that the algorithm converges in our setting.} We also compare and discuss the performance of the DNN models that are trained with zero-order optimization technique and that based on implicit function theorem (see, e.g.~\cite{donti2021dc3}) in Sec.~\ref{discuss.on.zero.order}.

\begin{algorithm}[t]
\caption{Proposed training algorithm}\label{alg:zoo.adam}
\LinesNumbered 
\KwIn{DNN with initial parameter $W_i^0$, $b_i^0$ for layer $i$, learning rate $\eta$, training epochs $T$, data-set $\mathcal{T}$}
\KwOut{Trained DNN with parameters $W_{i}^{T}$ and $b_{i}^{T}$}
$t=0$\\
\For{$t<T$}{
        Shuffle the training data set $\mathcal{T}$ \;
		\For{each batch $\mathcal{B} \subset \mathcal{T}$}{
            Compute the loss gradient via zero-order technique w.r.t DNN's output:\
		    $$
            \nabla \hat{\mathcal{L}}^t=\frac{1}{\mathrm{card}\left( \mathcal{B} \right)}\sum_{j\in \mathcal{B}}{\nabla \hat{\mathcal{L}}_{j}\left( \mathbf{s}_{pred,j} \right)}     
		    $$\\
		    Update $W_i$ and $b_i$ by back-propagation:\
		    \begin{eqnarray*}
            W_{i}^{t+1}&=&W_{i}^{t}-\eta \cdot \nabla \hat{\mathcal{L}}^{t}\left( W_{i}^{t} \right)\\
            b_{i}^{t+1}&=&b_{i}^{t}-\eta \cdot \nabla \hat{\mathcal{L}}^{t}\left(b_{i}^{t} \right)            
		    \end{eqnarray*}
		}
		 $t=t+1$
     }
\end{algorithm}
\vspace{-0.1in}
{\subsection{Feasibility Test and Recovery} \label{ssec:solve.post}
As discussed in Sec~\ref{ssec:Prediction and Reconstruction}, \textsf{DeepOPF} predicts a set of independent variables in the AC-OPF solution and reconstructs the remaining ones by solving AC power flow equations. As such, the AC power flow equality constraints are guaranteed to be satisfied. Meanwhile, due to the inherent DNN approximation error, the obtained AC-OPF solution may not satisfy the inequality constraints, i.e., generations/voltages/branch-flow limits, even with the penalty approach in place. 

Thus, after obtaining the solution, we first verify its feasibility by checking if it violates any inequality constraints. We output the solution if it passes the feasibility test. Otherwise, we run a conventional solver with the solution as a warm-start point to recover a feasible solution. If no feasible solution can be recovered, the particular load input is not supportable. 

We note that the feasibility-recovery step is usually faster than solving AC-OPF problems directly, especially when the warm-start point is close to the optimal~\cite{baker2019learning,diehl2019warm,dong2020smart}. Simulation results in Sec.~\ref{sec:simulations} show the obtained solutions from \textsf{DeepOPF} with the penalty approach are feasible for most of the time ($>99\%$) and do not need to go through feasibility recovery. The overall speedup performance is decent, with feasibility-recovery in place.
}

\vspace{-0.1in}
\section{Analysis of \textsf{DeepOPF}} \label{sec:OPF.theory_analysis}
In this section, we show that the AC-OPF load-to-solution mapping is continuous when the optimal solution is unique, adding new understanding to the AC-OPF problem. We then show that DNN can approximate such a mapping arbitrarily well as the number of neurons increases. 
\vspace{-0.1in}
\subsection{Understanding the AC-OPF Load-Solution Mapping}
Let $f^*(\cdot)$ denote the mapping from the load, denoted by $\mathbf{D}:=\left( P_{Di},Q_{Di},i\in \mathcal{N} \right)$, to the optimal solution of the AC-OPF problem, denoted by $\mathbf{X}^{*}:=\left( P^{*}_{Gi},Q^{*}_{Gi},|V^{*}_i|,\theta^{*}_i,i\in \mathcal{N} \right)$. Noted that the AC-OPF problem may have multiple optimal solutions for a given load input and in general $f^*(\cdot)$ may be a set of mappings and $\mathbf{X}^{*}$ may be a set of optimal solutions. Here we abuse the notation a bit to ease the following discussions. Before proceeding, we recall the definition of set of Lebesgue measure zero. 
\begin{defi}
    A set $S\subset \mathbb{R}^n$ has Lebesgue measure zero if for every $\epsilon > 0$, $S$ can be covered by a countable family of $n$-cubes, the sum of whose measures is less than $\epsilon$. 
\end{defi}
\noindent Taking 2-dimensional space as example, a set in the space has Lebesgue measure zero means its area is 0; thus a randomly-selected load input lies in the set with zero probability. The following theorem gives a useful observation on the smoothness of $f^*(\cdot)$ when the optimal solution is unique.

\begin{theorem} \label{thm3}
    Assumed the load domain is compact and $\mathbf{X}^{*}$ is unique for any given $\mathbf{D}$ in the load domain, $f^*(\cdot)$ is a continuous mapping. Further, the Hessian of $f^*(\cdot)$ exists everywhere over the load domain except for a set of Lebesgue measure zero.
\end{theorem}
\noindent We noted the unique optimal solution to AC-OPF problem is an empirical observation for networks, e.g., a distribution network with a tree topology, and there exist sufficient conditions to guarantee the uniqueness of the optimal solution (see, e.g.,~\cite{low2014convex2}). Theorem~\ref{thm3} says that, when the optimal solution is unique, the load-to-solution mapping $f^*(\cdot)$ is continuous everywhere and differentiable almost everywhere, e.g., it cannot be step or Dirichlet functions/mappings, adding new understanding to the AC-OPF load-to-solution mapping in the literature. This observation justifies the endeavors of leveraging DNN to learn the continuous load-to-solution mapping for a given power network. Specifically, there always exists a DNN-learned mapping $f\left(\cdot \right)$ whose approximation error to the continuous mapping $f^*\left( \cdot \right)$ can be arbitrarily small, as the number of neurons increases~\cite{HORNIK1991251}.
\vspace{-0.1in}
\subsection{DNN Approximation Error of Load-to-Solution Mapping}\label{ssec.dnn.approximation.error}
We further establish results for DNN approximation error to the load-to-solution mapping. Without loss of generality, we focus on the one-dimension output case in the follows discussion, i.e., $f^*(\cdot)$ is a scalar.\footnote{To extend the results for mapping with one-dimensional output to the mapping with multi-dimensional outputs, one can apply the results for one-dimensional output multiple times and combine them to get the desired result.} Suppose the input domain of load as the multi-dimensional unit hyper-cube, i.e., $\mathbf{D}\in \left[ \text{0,1} \right]^{\text{2*card}\left( \mathcal{N} \right)}$. We derive the following result by {extending} the analysis in~\cite{SafranS17} to the AC-OPF setting.
\begin{theorem}\label{lower.bound.ac.thm}
    Let $\mathbf{D}$ follow any continuous density function $p_{\mathbf{D}}$ over $\left[ \text{0,}\text{1} \right]^{2 \cdot\text{card}\left( \mathcal{N} \right)}$ that is lower-bounded by a positive constant. Let $\mathcal{K}_{m,L}$ be the class of all $f(\cdot)$ generated by a DNN with $L$ hidden layers and at most $m$ neurons per layer, and adopting ReLU as the activation function. \rev{Let $f^*(\cdot)$ be the unique load-solution mapping of an AC-OPF problem.} If the absolute value of its Hessian's eigenvalues is lower bounded by a positive constant in a connected subset of the input domain whose measure is lower bounded by a positive constant, \rev{we have: 
    \begin{equation}
    \underset{f\in \mathcal{K}_{m,L}}{\min}E_{_{\mathbf{D}\backsim u_{\mathbf{D}}}}\left[ \left( f^*\left( \mathbf{D} \right) -f\left( \mathbf{D} \right) \right) ^2 \right] \ge  \frac{c}{\left( 2m \right) ^{4L}},
    \label{lower.bound.ac}
    \end{equation}
    where the constant $c$ depends on the minimum of the density function and the minimum absolute value of the Hessian's eigenvalues.} 
\end{theorem}
\noindent Note that for AC-OPF problems, the Hessians of $f^*\left( \cdot \right)$ are usually not zero.~\footnote{The Hessians can be zeros almost everywhere for the piece-wise DC-OPF load-to-solution mapping~\cite{deepopf2}.} Thus it is not difficult for the ``bounded over a subset'' assumption on Hessian's eigenvalues to hold in practice, which essentially requires $f^*(\cdot)$ to have some ``curvature'' for the nonlinear AC-OPF problem. Theorem~\ref{lower.bound.ac.thm} implies an order-wise lower bound when applying DNN to approximate a load-to-solution mapping of any AC-OPF problem. The bound decreases exponentially in $L$ but polynomially in $m$. This highlights the benefits of using ``deep'' architecture in approximating the load-to-solution mapping of AC-OPF problems, similar to the observation for general functions in~\cite{SafranS17,LiangS17}. For mapping over general input domain with different ranges for individual loads, one can first scale up/down the loads to adjust their ranges to $[0,1]$ and then apply the lower bound result in the above theorem. \rev{The order-wise lower bound of approximation error in \eqref{lower.bound.ac} can be applied to uniform distribution (same as that in ~\cite{SafranS17}) and truncated normal distribution.
} For example, the constant involved in the right-hand-side of~\eqref{lower.bound.ac} is one for the case of uniform distribution. These results provide theoretical justification of using DNN to solve AC-OPF problem.

\begin{table}[!t]
	\centering
	\caption{Parameters settings.}
	\renewcommand{\arraystretch}{1.15}
	\begin{threeparttable}
		\begin{tabular}{c|c|c|c|c|c}
			\hline
			\tabincell{c}{\#Bus} & \tabincell{c}{\#P-V Bus} & \tabincell{c}{\#P-Q bus} &\tabincell{c}{\#Branch} &\tabincell{c}{\#Hidden\\layers} &\tabincell{c}{\#Neurons \\per layer} \\
			\hline
			30 & 5 & 24 &41 & 2 & 64/32\\
			\hline
			118 & 53 & 63 &231 & 2 & 256/128\\
			\hline
			300 & 68 & 231 & 411 & 2 & 512/256\\
			\hline
			2000 & 177 & 1822 & 3693 & 2 & 2048/1024\\
			\hline 			
		\end{tabular}
	\end{threeparttable}
	\label{table.parameter}
\end{table}

\begin{table*}[!t]
	\centering
	\caption{Performance comparisons for IEEE standard cases with uniformly-sampled load data ($\pm 10\%$ variation).}
	\renewcommand{\arraystretch}{1.15}
		\begin{threeparttable}[b]
			\begin{tabular}{c|c|c|c|c|c|c|c|c|c}
				\hline
				\multirow{2}{*}{\tabincell{c}{Test case}} &	\multicolumn{3}{c|}{\tabincell{c}{Feasibility rate (\%) \\before feasibility-$\text{recovery}^*$}} &
				\multicolumn{3}{c|}{\tabincell{c}{Average cost difference (\%)}} &
				\multicolumn{3}{c}{\tabincell{c}{Average speedup}} \\
				\cline{2-10}
                &\tabincell{c}{$\text{DNN-E}^{\dagger}$} & \tabincell{c}{DNN-W} & \tabincell{c}{\textsf{DeepOPF}} & \tabincell{c}{DNN-E} & \tabincell{c}{DNN-W} & \tabincell{c}{\textsf{DeepOPF}}&\tabincell{c}{DNN-E} & \tabincell{c}{DNN-W} & \tabincell{c}{\textsf{DeepOPF}}  \\
				\hline           
                \tabincell{c}{IEEE Case30}& 42 & 100 & 100& $<$0.1 & 0 & $<$0.1 & $\times$12 &$\times$1.1 &$\times$24  \\
    			\hline
                \tabincell{c}{IEEE Case118}& 22 &100 & 100& $<$0.1 & 0 & $<$0.1 &$\times$4.2 &$\times$1.3 &$\times$22  \\
    			\hline
                \tabincell{c}{IEEE Case300}& 21 & 100 & 99 & $<$0.1 & 0 & $<$0.1 & $\times$5.4 &$\times$1.8 &$\times$20  \\
    			\hline
                \tabincell{c}{IEEE Case2000}& 29 &100 & 99& $<$0.1 &0 & $<$0.1 & $\times$24 &$\times$1.0 &$\times$123 \\
				\hline
			\end{tabular}
		    \begin{tablenotes}
			\footnotesize
            \item[*] {Feasibility rates after feasibility-recovery are 100\% for \textsf{DNN-E} and \textsf{DeepOPF}.}
            \item[$\dagger$] {\rev{After obtaining the solution variables, \textsf{DNN-E} as in~\cite{zamzam2019learning} projects the infeasible ones (e.g., reactive power generation at P-V buses) onto the corresponding box constraints and re-solves the AC-PF equations one more time for the remaining variables. We note that such a post-processing process does not guarantee the final solutions respect the branch-flow limits, thus the low feasibility rates.}}            
		\end{tablenotes}
	\end{threeparttable}
	\label{table.10.sampled.load}
\end{table*}

\begin{table*}[!t]
	\centering
	\caption{Performance comparisons for IEEE standard cases with real-world load profiles with up to 40\% demand variation.}
	\renewcommand{\arraystretch}{1.2}
		\begin{threeparttable}[b]
			\begin{tabular}{c|c|c|c|c|c|c|c|c|c}
				\hline
				\multirow{2}{*}{\tabincell{c}{Test case}} &	\multicolumn{3}{c|}{\tabincell{c}{Feasibility rate (\%)\\ before feasibility-$\text{recovery}^*$}} &
				\multicolumn{3}{c|}{\tabincell{c}{Average cost difference (\%)}} &
				\multicolumn{3}{c}{\tabincell{c}{Average speedup}} \\
				\cline{2-10}
                &\tabincell{c}{DNN-E} & \tabincell{c}{DNN-W} & \tabincell{c}{\textsf{DeepOPF}} & \tabincell{c}{DNN-E} & \tabincell{c}{DNN-W} & \tabincell{c}{\textsf{DeepOPF}}&\tabincell{c}{DNN-E} & \tabincell{c}{DNN-W} & \tabincell{c}{\textsf{DeepOPF}}  \\
				\hline           
                \tabincell{c}{IEEE Case30}& 36 & 100 & 100& $<0.1$ & 0& $<0.1$ & $\times7.3$ &$\times1.0$ & $\times13$ \\
    			\hline
                \tabincell{c}{IEEE Case118}& 80 & 100 & 99 & $<0.1$ & 0 & $<0.1$ & $\times11$ &$\times1.1$ & $\times12$\\
    			\hline
                \tabincell{c}{IEEE Case300}& 49 & 100 & 100 & $<0.2$ & 0 & $<0.2$ & $\times16$ &$\times1.7$ & $\times33$ \\
    			\hline
                \tabincell{c}{IEEE Case2000}& 60 & 100 & 100 & $<0.2$ &0 & $<0.2$ & $\times44$ &$\times0.9$ & $\times70$ \\ 
				\hline
			\end{tabular}
		    \begin{tablenotes}
			\footnotesize
            \item[*] {Feasibility rates after feasibility-recovery are 100\% for \textsf{DNN-E} and \textsf{DeepOPF}.}
		    \end{tablenotes}
	\end{threeparttable}
	\label{table.real.world.load}
\end{table*}
\section{Numerical Experiments}\label{sec:simulations}

\subsection{Experiment Setup}\label{experiment.setup}
{We evaluate the performance of \textsf{DeepOPF} over the IEEE 30-/118-/300-bus~\cite{tpcwTrey30-118-300} test cases and a synthetic 2000-bus mesh power network~\cite{tpcwTrey2000} in the Power Grid Lib~\cite{pglib}.} Table~\ref{table.parameter} shows the related parameters for the test cases. As the Power Grid Lib only supports linear cost function for the IEEE 30-/118-/300-bus cases, we modify the cost parameters by applying the quadratic function parameters of the test cases from MATPOWER~\cite{zimmerman2011matpower} (ver. 7.0) while all other parameters remain unchanged. 

{We generate two datasets used for each test case in the simulation. The first dataset corresponds to the scenario where each load changes independently. In particular, the load data is sampled within $[90\%, 110\%]$ of the default load uniformly at random. Note that, theoretically, the DNN approach can be applied to learn the load-to-solution mapping of any load sampling range. The second dataset corresponds to the scenario in which loads at buses are correlated (e.g., the peak loads at most buses appear simultaneously). Specifically, similar to~\cite{tang2017real}, we generate the load data at each bus by multiplying the default value by an interpolated demand curve based on California’s daily net load from Dec. 1st to Dec. 8th, 2021, with a time granularity of 30 seconds. The generated real-time load data contains various correlated load demand variation up to 40\%. We then use the AC-OPF solutions obtained by Pypower~\cite{tpcwTrey1} as ground-truth. Finally, we generate uniformly-sampled and real-world datasets with 12,500 samples and use an 80\%/20\% split for training/test for each case.\footnote{{In our study, we adopt a common practice in recent works, e.g.,~\cite{chatzos2020high,singh2021learning}, to select the size of training set and test set. From our simulation results, this amount of data is sufficient to obtain strong performance for the case studies. We leave the problem of determining the minimum number of samples needed to train the DNN with strong performance as a future direction.}}
}

We design the DNN model based on the Pytorch platform and integrate the zero-order optimization technique with the widely-used Adam~\cite{adam} algorithm for training. The training epoch and the batch size for all test cases are 200 and 32, respectively. We set the weighting factors in the loss function in \eqref{total.loss} to be $w_1 = 1, w_2 = 0.1$ based on empirical experience. 
Table~\ref{table.parameter} also shows other related parameters of the DNN model, e.g., the number of hidden layers and the number of neurons in each layer.\footnote{{We note that Theorem~\ref{lower.bound.ac.thm} in Sec.~\ref{sec:OPF.theory_analysis} shows the benefits of increasing the size of DNN in mapping approximation. However, the over-fitting issue may appear in practice for larger DNNs, resulting in performance degeneration. We follow a common practice to determine the DNN structure and size by educated guesses and iterative tuning for different test cases.}}

We conduct simulations in CentOS 7.6 with quad-core (i7-3770@3.40G Hz) CPU and 16GB RAM. {In addition to the state-of-the-art iterative solver (Pypower), we compare \textsf{DeepOPF} with the following DNN schemes:
\begin{itemize}
\item \textsf{DNN-E}: A method adapted from the one used in~\cite{zamzam2019learning}. It trains a DNN to learn the load-solution mapping for solving AC-OPF problems without the penalty approach.
\item \textsf{DNN-W}: A method adapted from the one used in~\cite{dong2020smart}. It trains a DNN to predict the primal variables as the warm-start points to the conventional solver.
\end{itemize}
For a fair comparison, the training/testing samples are all set the same as \textsf{DeepOPF}, and the feasibility-recovery procedure is applied in case of in-feasibility.} The evaluation metrics are the following (averaged over 2,500 test instances): (i) {Feasibility rate}: The percentage of the obtained feasible solution before feasibility-recovery procedure. A solution is regarded feasible only if it satisfies all AC-OPF constraints.\footnote{In~\cite{zamzam2019learning,chatzos2020high}, the feasibility is evaluated by the constraint violation rate average over all test instances. This, however, does not measure the fraction of the instances with all constraints satisfied, which is our feasibility definition.} (ii) {Cost}: The corresponding power generation cost difference as compared to the Pypower. 
(iii) {Speedup}: The average running-time ratios of the Pypower to that of the DNN schemes taking the feasibility-recovery procedure are computed as speedup. Note that the speedup is the average of ratios, and it is different from the ratio of the average running times between the Pypower and DNN schemes. 
{\subsection{Performance Evaluation under Test Datasets}
We show the simulation results of the proposed approach for the test cases with the synthetic and real-world load profiles in Tables~\ref{table.10.sampled.load} and ~\ref{table.real.world.load}. We have the following observations. First, substantial improvements in the feasibility can be achieved when applying the penalty approach. The feasibility rate increases up to~80\% as compared to \textsf{DNN-E}.\footnote{{Note that in \textsf{DNN-E}, the power-flow equality constraints are always satisfied. But the reactive power and line limit constraints may be violated as they are not considered during the DNN training.}} Overall, \textsf{DeepOPF} obtains $>99\%$ feasibility rate before the feasibility-recovery procedure, demonstrating the usefulness of the penalty approach. Second, \textsf{DeepOPF} achieves an average optimality loss of less than $0.2\%$ and a speedup performance of $\times$123 (two orders of magnitude), as compared to conventional AC-OPF solver on two datasets. It demonstrates the effectiveness of \textsf{DeepOPF} in dealing with both small-/large-variation load profiles. We also compare the DNN-predict solution with the ground-truths for IEEE Case118 for a specific test instance in Fig.~\ref{fig2}. The results show that \textsf{DeepOPF} can achieve desirable prediction results and thus suggest minor optimality loss. In addition, \textsf{DeepOPF} achieves decent speedup performance, which has included time used by the feasibility-recovery procedure. The speedup performance also demonstrates the scalability of the proposed approach and its potential for solving larger-scale AC-OPF problems. Third, as compared to the hybrid scheme (\textsf{DNN-W}) using DNN to predict the warm-start points, \textsf{DeepOPF} achieves noticeably better speedup performance and similar feasibility and optimality performance. These results demonstrates effectiveness of employing DNNs to solve AC-OPF problems directly.
}


\subsection{Performance with Different Penalty Weighting Factors}
We also evaluate the performance of \textsf{DeepOPF} with different weighting factors in the loss function for training. Specifically, we use IEEE Case118 {with synthetic load profiles} and choose two variants of the weighting factors as follows:
\begin{itemize}
\item \textsf{Weights-V1}: $w_1=1$ and $w_2=1$.
\item \textsf{Weights-V2}: $w_1=1$ and $w_2=0.1$.
\end{itemize}
They assign different priorities in the prediction error and the penalty term to pursue optimality and preserve feasibility. We compare two variants' results in terms of feasibility rate, speedup, and the average optimality gap to the conventional solver on the test data set. As seen in Table~\ref{table5}, although adding penalty improves the speedup of the approach as it involves less in-feasible instances, a larger value of $w_2$ may cause a higher optimality difference as the DNN model pays more attention to reduce the penalty during the training, affecting prediction performance and the final optimality difference. \rev{In practice, one can start with an initial value and then tune the value towards satisfactory training performance. We leave setting weighting factors systematic as a future direction.}

\begin{figure*}[!t]
	\centering
	\subfigure [] {\includegraphics[width = 0.45\textwidth]{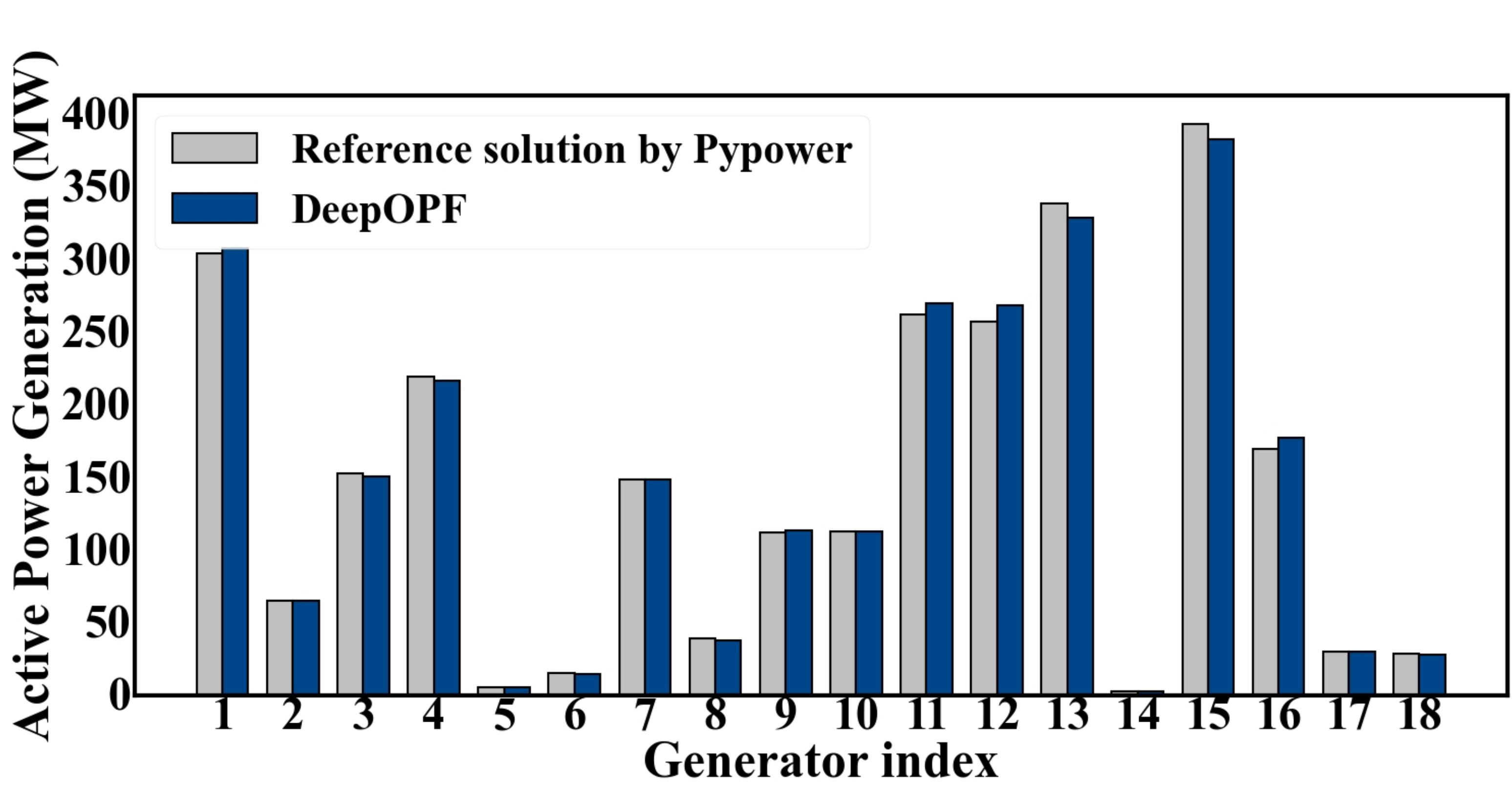}}
	\subfigure [] {\includegraphics[width = 0.45\textwidth]{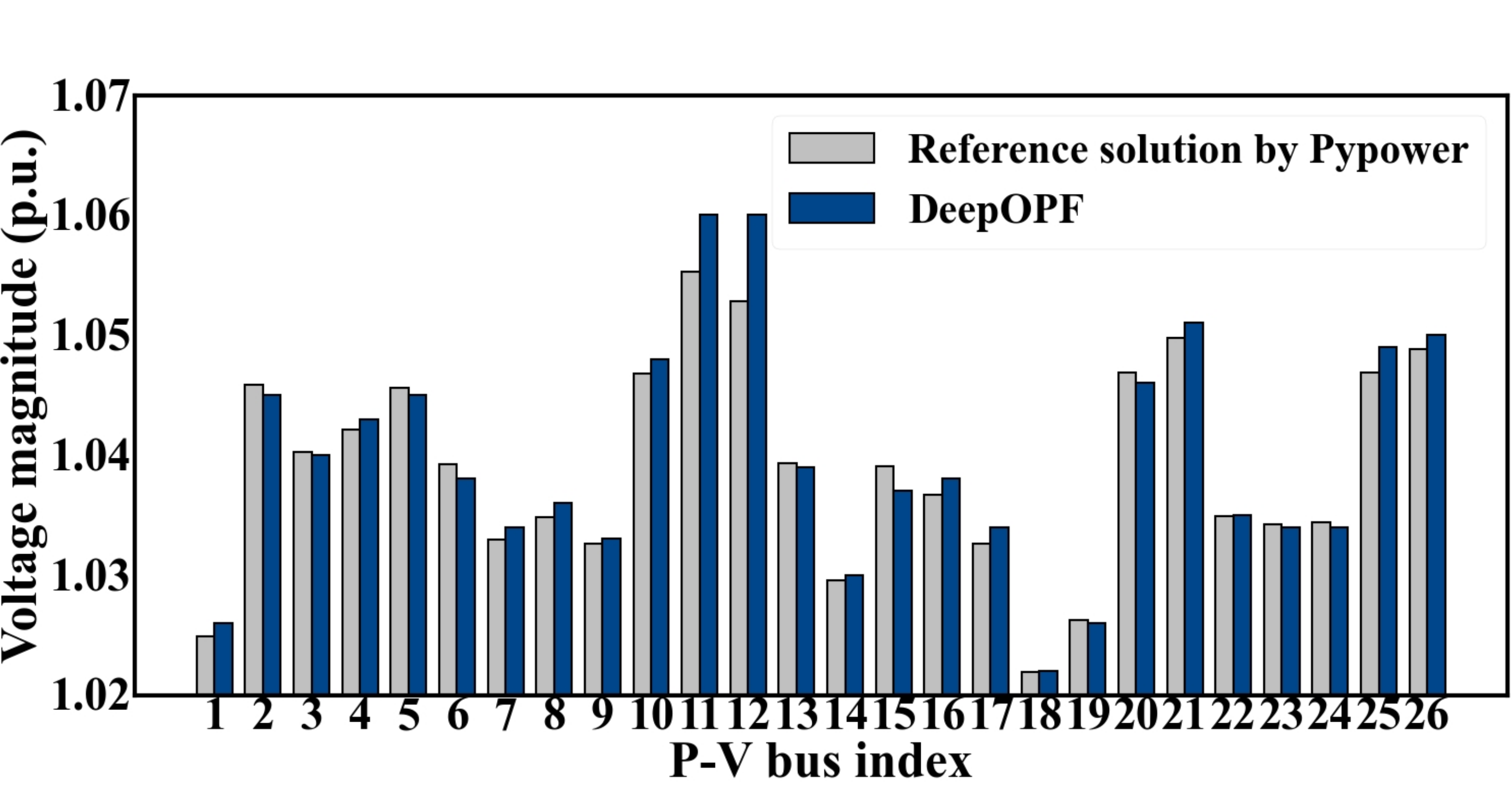}}
	\captionsetup{font={small}}
	\caption{{Comparisons of the \textsf{DeepOPF} solutions and the Pypower solutions for the IEEE Case118 test case. There are 53 P-V buses in total, and 18 of them are associated with active power generation. (a) The differences on active power generations on the 18 PV buses with active power generation. (b) The differences on voltage magnitudes at the first 26 P-V buses.}}
	\label{fig2}
\end{figure*}

\subsection{Performance \hspace{-0.05em}with \hspace{-0.05em}Different \hspace{-0.05em}Gradient Computation Methods}\label{discuss.on.zero.order}
We carry out experiments under the same setting to compare the optimality loss and feasibility of two \textsf{DeepOPF} variants on IEEE case30 and case118 test cases {with the synthetic load profiles}:
\begin{itemize}
    \item \textsf{DeepOPF-ZO}: the \textsf{DeepOPF} with zero-order technique for estimating the penalty gradient.
    \item  \textsf{DeepOPF-IF}: {the \textsf{DeepOPF} scheme by utilizing implicit function theorem, to compute the penalty gradient. We noted that this is the method adapted from the one used in~\cite{donti2021dc3}.} The idea is to first compute the partial derivatives of power flows w.r.t $x_{pred}$ and $x_{rec}$ and then obtain the gradient of $x_{rec}$ w.r.t $x_{pred}$ and the $\nabla\mathcal{L}_{pen}\left(s_{pred} \right)$ by using the chain rule.
\end{itemize}
The results are shown in Table.~\ref{table6}. Interestingly, we observe in the simulations that \textsf{DeepOPF-ZO} achieves the same performance as \textsf{DeepOPF-IF}, except for a better feasibility rate. This indicates that \textsf{DeepOPF-ZO} is able to train a DNN with similar, sometimes better, performance than \textsf{DeepOPF-IF}. The explanation for this observation may be that the estimated gradient (with noise) could help the DNN model to escape the bad local minimum during the training process and achieve better performance. Similar observations are also reported in~\cite{neelakantan2015adding} for training DNN in other problem domains.

In general, both schemes have their own merits. \textsf{DeepOPF-IF} can compute the exact gradient to guide the training, given a well-characterized and computationally efficient model. In practice, however, such a model may be difficult to obtain due to sophisticated equipment operation mechanisms in the power system. For example, for the AC-OPF problem with wind turbine controls for increasing power capture, it is difficult to model the relationship between the control variable and wind-turbine output due to the complex control mechanism~\cite{marden2013model}. In practice, the wind farm's output evaluation proceeds through simulation/measurement apparatus, and no mathematical model exists for computing penalty gradients directly. For such problems, it may be more convenient to use \textsf{DeepOPF-ZO} instead of \textsf{DeepOPF-IF}.

\begin{table}[!t]
\caption{Performance comparisons of weighting-factor variants.}
	\renewcommand{\arraystretch}{1.2}
	\centering
	\begin{tabular}{c|c|c|c}
		\hline
        \multicolumn{1}{c|}{Weight setting} & \tabincell{c}{Feasibility rate (\%) \\before feasibility-recovery} & \tabincell{c}{Cost\\diff. (\%)} &\tabincell{c}{Avg.\\ speedup} \\
        \hline
        \tabincell{c}{\textsf{Weights}-V1} &100 & 1.5& $\times$ 22\\
        \hline
        \tabincell{c}{\textsf{Weights}-V2} &100 & $<$0.1 & $\times$ 22 \\
		\hline
		\end{tabular}
	\label{table5}
\end{table}

\begin{table}[!t]
	\centering
	\caption{Performance comparison of \textsf{DeepOPF} variants.}
	\renewcommand{\arraystretch}{1.2}
	\begin{threeparttable}	
	\begin{tabular}{c|c|c|c|c}
	\hline
	\tabincell{c}{Test\\ case}&				
	\tabincell{c}{Variants}&
	\tabincell{c}{Feasibility rate (\%) \\before\\ feasibility-recovery}&
    \tabincell{c}{Cost \\diff(\%)} &
	\tabincell{c}{Avg. \\speedup} \\
	\hline           
	\multirow{2}{*}{\tabincell{c}{Case-30}} &\tabincell{c}{\textsf{DeepOPF-ZO}} &100&$<$0.1&$\times$24\\
    \cline{2-5}
    & \tabincell{c}{\textsf{DeepOPF-IF}} &100&$<$0.1&$\times$24\\
    \hline
    \multirow{2}{*}{\tabincell{c}{Case-118}} &\tabincell{c}{\textsf{DeepOPF-ZO}} &100&$<$0.1&$\times$22\\
    \cline{2-5}
    & \tabincell{c}{\textsf{DeepOPF-IF}} &95&$<$0.1&$\times$21\\
    \hline
	\end{tabular}
	\end{threeparttable}	
	\label{table6}
\end{table}

\vspace{-0.1in}
\section{Conclusion and Future Directions}\label{sec:conclusion}
We develop a feasibility-optimized DNN for solving AC-OPF problems. To ensure that the power-flow balance constraints are satisfied, \textsf{DeepOPF} first predicts a set of independent variables and then reconstructs the remaining variables by solving the AC power flow equations. We also adopt a penalty approach in the DNN training to respect the inequality constraints. We further apply a zero-order optimization-based training algorithm to compute the penalty gradient efficiently. Simulation results on IEEE 30/118/300-bus and a synthetic 2000-bus test cases show the effectiveness of the penalty approach and that \textsf{DeepOPF} speeds up the computing time by two-orders of magnitude as compared to conventional optimization-based solvers with minor cost difference. Empirical results also reveal the advantage of zero-order technique on training DNN model to achieve better performance. \textsf{DeepOPF} develops one DNN model for solving AC-OPF problems with varying load inputs but fixed power network topology and parameters (e.g., line impedance) and other physical/operational constraints. Thus, upon contingencies, e.g., line outage and transmission switching, one may need to load a pre-trained DNN model to solve AC-OPF problems with revised topology, line parameters, or constraints. It is of great interest to explore DNN designs that are robust to changes in system \rev{topologies} or parameters. \rev{Another interesting future directions is to extend \textsf{DeepOPF} to AC-OPF problems with multiple load-solution mappings~\cite{pan2022deepopf}.}
\section*{Acknowledgement}
We thank Titing Cui, Wanjun Huang, Qiulin Lin, and Andreas Venzke for the discussions related to this study. \rev{We would like to thank Ying Guo and the National Supercomputer Center in Jinan for providing GPU/CPU computing resources. We would alos like to thank the anonymous reviewers for careful reading and the helpful comments.}



\bibliographystyle{IEEEtran}
\bibliography{IEEEabrv,ref}


\appendix
\begin{appendices}


\section{Proof of Theorem~\ref{thm3}} \label{thm3_proof}
\begin{proof}
AC-OPF problem can be regarded as the following nonlinear programming problem:
\begin{eqnarray}
    &\min_{\mathbf{X}}& \,\,c\left( \mathbf{X,D} \right) \nonumber\\
    &\mathrm{s.t.\;\;}&g_i\left( \mathbf{X,D} \right) \le 0,\,\,i \in \mathcal{M},
    \label{ac.opf.nlp}
\end{eqnarray}
\noindent where the load input and variables for AC-OPF problem are denoted by $\mathbf{D}:=\left( P_{Di},Q_{Di},i\in \mathcal{N} \right)$ and  $\mathbf{X}:=\left( P_{Gi},Q_{Gi},|V_i|,\theta_i,i\in \mathcal{N} \right)$, respectively. Functions $c\left(\cdot \right)$ and $g_i\left( \cdot \right)$ are continuous w.r.t $\mathbf{D}$ and $\mathbf{X}$. $\mathcal{M}$ is the set of constraints. Under the setting of the theorem, we denote the unique optimal solution by $\mathbf{X}^{*}$ and the load-to-solution mapping by $f^*(\cdot)$, i.e., $\mathbf{X}^{*}=f^*\left( \mathbf{D} \right)$. As the variables of AC-OPF problem usually have bounded constraints, $f^*(\cdot)$ is supposed to be bounded without loss of generality. We further assume the input domain of $\mathbf{D}$ (say $\mathbf{\varOmega_D}$) is compact. Before proceeding, we recall the definition of sequence and sub-sequence.


\begin{defi}
    Let $\mathbb{N}_+$ denote the positive integer set. A sequence is a function from a subset of integers (e.g., $\mathbb{N}_+$) to another set (e.g., $\mathbb{R}^m, m \in \mathbb{N}_+$). A sub-sequence is a sequence derived from a sequence by deleting some/no elements without changing the order of the remaining elements.
\end{defi}
\noindent We then define the convergence of a sequence.
\begin{defi}
    A sequence $\left\{ \boldsymbol{x}_n|\boldsymbol{x}_n\in \mathbb{R} ^m \right\} _{n=1}^{\infty}$ is said to converge to $\boldsymbol{a}$ ($\boldsymbol{a} \in \mathbb{R}^m$) if for any $\xi>0$, there exists $N'$ such that, $\left\| \boldsymbol{x}_n-\boldsymbol{a} \right\| _2<\xi$ holds for any $n>N'$; $\boldsymbol{a}$ is called the limit of sequence $\left\{ \boldsymbol{x}_n \right\} _{n=1}^{\infty}$, written as $\underset{n\rightarrow \infty}{\lim}\boldsymbol{x}_n=\boldsymbol{a}$.
    \label{defi6}
\end{defi}
\noindent By Definition~\ref{defi6}, we have the definition of continuous function in terms of sequences' limits. 
\begin{defi}
    A function $f(\cdot)$ defined on a input domain $\mathbf{\varOmega}$ ($\mathbf{\varOmega} \subseteq \mathbb{R}^m$) is continuous if for any sequence generated on the input domain, i.e., $\left\{ \boldsymbol{x}_n|\boldsymbol{x}_n\in \mathbf{\varOmega } \right\} _{n=1}^{\infty}$, converges to $\boldsymbol{c}$, the corresponding sequence $\left\{ f(\boldsymbol{x}_n) \right\} _{n=1}^{\infty}$ converges to $f(\boldsymbol{c})$.
    \label{defi7}
\end{defi}
\noindent As we aims to show the discontinuity of $f^*(\cdot)$, we further derive the following lemma to characterize a discontinuous bounded function.
\begin{lemma}
    If a bounded function $f(\cdot)$ defined on a input domain $\mathbf{\varOmega}$ ($\mathbf{\varOmega} \subseteq \mathbb{R}^m$) is discontinuous, there exists a sequence generated on the input domain, i.e., $\left\{ \boldsymbol{x}_n|\boldsymbol{x}_n\in \mathbf{\varOmega } \right\} _{n=1}^{\infty}$, converges to $\boldsymbol{c}$ and the corresponding sequence $\left\{ f(\boldsymbol{x}_{n}) \right\}_{n=1}^{\infty}$ converges to a different value other than $f(\boldsymbol{c})$.
    \label{lemma7}
\end{lemma}
\begin{proof}
By Definition~\ref{defi7}, we can obtain the following two cases to characterize the general discontinuous function $f(\cdot)$ defined on a input domain $\mathbf{\varOmega}$.
\begin{itemize}
\item \textbf{Case 1}: There exist a sequence $\left\{ \boldsymbol{x}_n|\boldsymbol{x}_n\in \mathbf{\varOmega } \right\} _{n=1}^{\infty}$ converges to $\boldsymbol{c}$ but the corresponding sequence $\left\{ f(\boldsymbol{x}_n) \right\} _{n=1}^{\infty}$ does not converge, i.e., $\underset{n\rightarrow \infty}{\lim}f\left( \boldsymbol{x}_n \right)$ does not exist;
\item \textbf{Case 2}: There exist a sequence $\left\{ \boldsymbol{x}_n|\boldsymbol{x}_n\in \mathbf{\varOmega } \right\} _{n=1}^{\infty}$ converges to $\boldsymbol{c}$ but the corresponding sequence $\left\{ f(\boldsymbol{x}_n) \right\} _{n=1}^{\infty}$ converges to a different value other than $f(\boldsymbol{c})$. In such case, Lemma~\ref{lemma7} holds trivially.
\end{itemize}
We then consider \textbf{Case 1} when $f(\cdot)$ is a bounded function. Recall that sequence $\left\{ f(\boldsymbol{x}_n) \right\} _{n=1}^{\infty}$ does not converge in such case. Since $f(\cdot)$ is bounded, by Bolzano-Weierstrass theorem~\cite{bartle2000introduction}, there exists a convergent sub-sequence, i.e., $\left\{ f(\tilde{\boldsymbol{x}}_m) \right\} _{m=1}^{\infty} \subseteq \left\{ f(\boldsymbol{x}_n) \right\} _{n=1}^{\infty}$ and $\underset{m\rightarrow \infty}{\lim}f(\tilde{\boldsymbol{x}}_m)$ exists (suppose the limit is $b$). As the sequence $\left\{ f(\boldsymbol{x}_n) \right\} _{n=1}^{\infty}$ does not converge, by definition, there exists $\xi>0$ such that for any $N'>0$, there exists $n>N'$ and $\left\| f(\boldsymbol{x}_k)-b \right\| _2>\xi$. Thus, given $\xi$, there exists a sub sequence $\left\{ f(\boldsymbol{x}_k)|\left\| f(\boldsymbol{x}_k)-b \right\| _2>\xi \right\} _{k=N'}^{\infty}$. 
By Bolzano-Weierstrass theorem~\cite{bartle2000introduction}, there exists another convergent sub-sequence within, denoted as $\left\{ f(\boldsymbol{\hat{x}}_m) \right\} _{m=1}^{\infty}\subseteq \left\{ f(\boldsymbol{x}_k)|\left\| f(\boldsymbol{x}_k)-b \right\| _2>\xi \right\} _{k=N'}^{\infty}$, that does not converge to $b$. To this end, if sequence $\left\{ \boldsymbol{x}_n|\boldsymbol{x}_n\in \mathbf{\varOmega } \right\} _{n=1}^{\infty}$ converges to $\boldsymbol{c}$, we can observe two sub-sequences within $\left\{ f(\boldsymbol{x}_n) \right\} _{n=1}^{\infty}$ that converge to different values; one of the two values must not equal to $f(\boldsymbol{c})$. Thus, we finish our proof by showing Lemma~\ref{lemma7} also holds in \textbf{Case 1}.
\end{proof}

Lemma~\ref{lemma7} provides the existence of convergent sequences for discontinuous bounded function, based on which we can use prove by contradiction to show the continuity of $f^*(\cdot)$. Denote $\mathbf{X}_k = f^*(\mathbf{D}_k)$, $k$ is natural number. Assuming $f^*(\cdot)$ is discontinuous at point $\mathbf{D}_0$, it means there exists a sequence $\left\{ \mathbf{D}_k,\mathbf{X}_k  |\mathbf{D}_k\in \mathbf{\varOmega }_{\mathbf{D}} \right\} _{k=1}^{\infty}$ such that $\underset{k\rightarrow \infty}{\lim}\mathbf{D}_k=\mathbf{D}_0$ and $\underset{k\rightarrow \infty}{\lim}\mathbf{X}_k =\mathbf{X}^{'} \ne \mathbf{X}_0$. Our main idea is to show $\mathbf{X}^{'}$ is also the optimal solution at the point $\mathbf{D}_0$, and then derive a contradiction with the uniqueness of the optimal solution. 


We first show the \textit{feasibility} of $\mathbf{X}^{'}$ at point $\mathbf{D}_0$, i.e., $g_i\left( \mathbf{X}^{'},\mathbf{D}_0 \right) \le 0, i \in \mathcal{M}$. Recall that we have $g_i\left( \mathbf{X}_k,\mathbf{D}_k \right) \le 0, i \in \mathcal{M}$, $k$ is a natural number. Let $\mathbf{Z}$ denote the concatenation of $\mathbf{X}$ and $\mathbf{D}$, i.e.,$\mathbf{Z}(\mathbf{X},\mathbf{D})=[\mathbf{X};\mathbf{D}]$. By continuity, for each $g_i\left(\cdot \right), i \in \mathcal{M}$, we have:
\begin{eqnarray}
\nonumber
&\forall \xi>0, \exists \delta_i>0,& \mathrm{s.t.} \left\|\mathbf{Z}(\mathbf{X},\mathbf{D})-\mathbf{Z}(\mathbf{\tilde{X}},\mathbf{\tilde{D}}) \right\| _2<\delta_i,\\
\label{eq20}
&&\left| g_i\left( \mathbf{X},\mathbf{D} \right) -g_i\left( \mathbf{\tilde{X}},\mathbf{\tilde{D}} \right) \right|<\mathrm{\xi}.
\end{eqnarray}
Since $\underset{k\rightarrow \infty}{\lim}\mathbf{D}_k=\mathbf{D}_0$ and $\underset{k\rightarrow \infty}{\lim} \mathbf{X}_k =\mathbf{X}^{'}$. By Definition~\ref{defi6} and~\eqref{eq20}, we have:
\begin{eqnarray}
\label{eq21}
&&\exists K>0, \mathrm{\ s.t.\ }\text{for}\ k>K, \nonumber\\
&&\left\| \mathbf{D}_k-\mathbf{D}_0 \right\| _2<\frac{\delta_i}{2}, \left\| \mathbf{X}_k-\mathbf{X}^{'} \right\| _2<\frac{\delta_i}{2},\\
&&\left| g_i\left( \mathbf{X}^{'},\mathbf{D}_0 \right) -g_i\left( \mathbf{X}_k,\mathbf{D}_k \right) \right|<\mathrm{\xi}, i \in \mathcal{M}.\nonumber
\end{eqnarray}

We then prove $g_i\left( \mathbf{X}^{'},\mathbf{D}_0 \right) \le 0, i \in \mathcal{M}$ by contradiction. Suppose there exists $i \in \mathcal{M}$ such that $g_i\left( \mathbf{X}^{'},\mathbf{D}_0 \right) >0$. We then can let $\xi=\frac{g_i\left( \mathbf{X}^{'},\mathbf{D}_0 \right)}{2}$ and leverage~\eqref{eq20} to obtain: 
\begin{equation}
    \exists \mathbf{X}_k,\mathbf{D}_k \,\,\mathrm{s}.\mathrm{t}. g_i\left( \mathbf{X}_k,\mathbf{D}_k \right) >\mathrm{\xi}>0,
\end{equation}
\noindent which is contradicted with $g_i\left( \mathbf{X}_k,\mathbf{D}_k \right) \le 0, i \in \mathcal{M}$. Thus, we have $g_i\left( \mathbf{X}^{'},\mathbf{D}_0 \right) \le 0, i \in \mathcal{M}$. 

We then show the \textit{optimality} of $\mathbf{X}^{'}$ at point $\mathbf{D}_0$, i.e., $\mathrm{c}\left( \mathbf{X}^{'},\mathbf{D}_0 \right)=\mathrm{c}\left( \mathbf{X}_0,\mathbf{D}_0 \right)$. Specifically, we need to prove: 
\begin{eqnarray}
\label{eq1}
\underset{k\rightarrow \infty}{\lim}\left| \mathrm{c}\left( \mathbf{X}_{k},\mathbf{D}_{k} \right) -\mathrm{c}\left( \mathbf{X}_0,\mathbf{D}_0 \right) \right|=0,\\
\label{eq2}
\underset{k\rightarrow \infty}{\lim}\left| \mathrm{c}\left( \mathbf{X}_k,\mathbf{D}_{k} \right) -\mathrm{c}\left( \mathbf{X}^{'},\mathbf{D}_0 \right) \right|=0.
\end{eqnarray}

To prove~\eqref{eq1}, we consider following two cases.~\footnote{For the sequence $\left\{ (\mathbf{X}_k,\mathbf{D}_k) \right\} _{k=1}^{\infty}$ in which $c\left( \mathbf{X}_{k},\mathbf{D}_{k} \right)$ is not larger or smaller than $\mathrm{c}\left( \mathbf{X}_0,\mathbf{D}_0 \right)$ identically, we can divide the sequence into two complementary sub-sequences: $\left\{ (\mathbf{X}_k,\mathbf{D}_k)|c\left( \mathbf{X}_k,\mathbf{D}_k \right) >\mathrm{c}\left( \mathbf{X}_0,\mathbf{D}_0 \right) \right\} _{k=1}^{\infty}
$ and $\left\{ (\mathbf{X}_k,\mathbf{D}_k)|c\left( \mathbf{X}_k,\mathbf{D}_k \right) \leqslant \mathrm{c}\left( \mathbf{X}_0,\mathbf{D}_0 \right) \right\} _{k=1}^{\infty}
$. We then can adopt similar idea to show each sub-sequence can converge to the same value, and thus the original sequence also converge to this value.}
\begin{itemize}
\item $c\left( \mathbf{X}_{k},\mathbf{D}_{k} \right) >\mathrm{c}\left( \mathbf{X}_0,\mathbf{D}_0 \right)$. In this case, we can first construct a sequence as follows. Let:
\begin{eqnarray}
\mathbf{\hat{X}}_{k}=&\mathrm{argmin}_{\mathbf{Y}}& \left\| \mathbf{Y}-\mathbf{X}_0 \right\|^{2}_2 \nonumber\\
&\mathrm{s.t.\;\;}&g_i\left( \mathbf{Y},\mathbf{D}_k \right) \le 0,\,\,i \in \mathcal{M}.
\label{eq24}
\end{eqnarray}
We following show $\mathbf{\hat{X}}_{k}$ converges to $\mathbf{X}_0$, by  considering three sub-cases: 
\begin{itemize}
\item \textbf{sub-cases 1} $\left\{ \mathbf{X}:g_i\left( \mathbf{X},\mathbf{D}_0 \right) <0,i\in \mathcal{M} \right\} $ is not an empty set and $g_i\left( \mathbf{X}_0,\mathbf{D}_0 \right) < 0, i \in \mathcal{M}$.~\footnote{For AC-OPF problems, we can divide the variables into independent and dependent variables. The equality constraints determine their relationship. Thus, we can remove the equality constraints by representing dependent variables by independent variables. Therefore, the considered problem is only with inequality constraints.} Similar to~\eqref{eq20}, we have:
\begin{eqnarray}
\nonumber
\forall \xi>0, \exists \delta_i>0, \mathrm{s.t.} \left\|\mathbf{Z}(\mathbf{X}_0,\mathbf{D}_0)-\mathbf{Z}(\mathbf{X}_k,\mathbf{D}_k) \right\| _2<\delta_i,&&\\
\nonumber
\left| g_i\left( \mathbf{X}_0,\mathbf{D}_0 \right) -g_i\left( \mathbf{X}_k,\mathbf{D}_k\right) \right|<\xi.&&
\label{eq26}
\end{eqnarray}
Let $g_i\left( \mathbf{X}_0,\mathbf{D}_0 \right) =\epsilon <0,\xi =\frac{\left| \epsilon \right|}{2}$. By above inequality, we can have:
\begin{eqnarray}
\label{eq27}
&&\exists \delta_i>0, \mathrm{\ s.t.\ }\left\| \mathbf{D}_k-\mathbf{D}_0 \right\| _2<\frac{\delta_i}{2},\\
&&\left| g_i\left( \mathbf{X}_0,\mathbf{D}_0 \right) -g_i\left( \mathbf{X}_0,\mathbf{D}_k \right) \right|<\mathrm{\xi}, i \in \mathcal{M}.\nonumber
\end{eqnarray}
Accordingly, we can have $g_i\left( \mathbf{X}_0,\mathbf{D}_k \right) <g_i\left( \mathbf{X}_0,\mathbf{D}_0 \right) +\xi <0$. Thus, it means $\mathbf{X}_{0}$ is a feasible solution at point $\mathbf{D}_k$ and we can have:$ \underset{k\rightarrow \infty}{\lim}\mathbf{\hat{X}}_{k}=\mathbf{X}_0$.

\item \textbf{sub-cases 2} $\left\{ \mathbf{X}:g_i\left( \mathbf{X},\mathbf{D}_0 \right) <0,i\in \mathcal{M} \right\} $ is not an empty set but $g_i\left( \mathbf{X}_0,\mathbf{D}_0 \right) = 0, i \in \mathcal{M}$. In this case, we can adopt similar ideas in \textbf{sub-cases 1} to show $\mathbf{\hat{X}}_{k}$ converges to $\mathbf{X}_0$.

\item \textbf{sub-cases 3} $\left\{ \mathbf{X}:g_i\left( \mathbf{X},\mathbf{D}_0 \right) <0,i\in \mathcal{M} \right\} $ is an empty set, it means the feasible solution on $\mathbf{D}_0$ must satisfy the equality constraints: $g_i\left( \mathbf{X},\mathbf{D}_0 \right) = 0, i \in \mathcal{M}$. In such case, 
the relationship between $\mathbf{X}$ and $\mathbf{D}_0$ is determined by the equality constraints. We then can remove the equality constraints by representing $\mathbf{X}_k$ with $\mathbf{D}_k$. Since $\mathbf{D}_k$ converges to $\mathbf{D}_0$ and $g_i\left( \mathbf{X}_0,\mathbf{D}_0 \right) = 0, i \in \mathcal{M}$, we can obtain $\mathbf{\hat{X}}_{k}$ converges to $\mathbf{X}_0$.
\end{itemize}
In summary, we show $\underset{k\rightarrow \infty}{\lim}\mathbf{\hat{X}}_{\mathrm{k}}=\mathbf{X}_0$. Since $\mathbf{X}_{k}$ is the unique optimal solution at point $\mathbf{D}_{k}$, we have $\mathrm{c}\left( \mathbf{X}_k,\mathbf{D}_k \right) \le \mathrm{c}\left( \mathbf{\hat{X}}_{\mathrm{k}},\mathbf{D}_{k} \right)$. Accordingly, we have:
\begin{eqnarray}
&&\underset{k\rightarrow \infty}{\lim}\left| \mathrm{c}\left( \mathbf{X}_{k},\mathbf{D}_{k} \right) -\mathrm{c}\left( \mathbf{X}_0,\mathbf{D}_0 \right) \right|\\\nonumber
&\le&\underset{k\rightarrow \infty}{\lim}\left| \mathrm{c}\left( \mathbf{\hat{X}}_{\mathrm{k}},\mathbf{D}_{k} \right) -\mathrm{c}\left( \mathbf{X}_0,\mathbf{D}_0 \right) \right|\\
&=&0.\nonumber
\end{eqnarray}
\item  $c\left( \mathbf{X}_{k},\mathbf{D}_{k} \right) \le \mathrm{c}\left( \mathbf{X}_0,\mathbf{D}_0 \right)$. In this case, we can have:
\begin{equation}
\underset{k\rightarrow \infty}{\lim}\mathrm{c}\left( \mathbf{X}_{k}, \mathbf{D}_{k} \right) \le \mathrm{c}\left( \mathbf{X}_0,\mathbf{D}_0 \right).
\end{equation}
\noindent As $\underset{k\rightarrow \infty}{\lim}\mathbf{D}_k=\mathbf{D}_0$ and $\mathbf{X}_0$ is the unique optimal solution at point $\mathbf{D}_0$, we have $\underset{k\rightarrow \infty}{\lim}\mathrm{c}\left( \mathbf{X}_{k}, \mathbf{D}_{k} \right) = \mathrm{c}\left( \mathbf{X}_0,\mathbf{D}_0 \right)$
\end{itemize}
Combing the results for above two cases, we show~\eqref{eq1} holds. We below prove~\eqref{eq2}. Since $\underset{k\rightarrow \infty}{\lim}\mathbf{D}_k=\mathbf{D}_0$ and $\underset{k\rightarrow \infty}{\lim}f^*\left( \mathbf{D}_k \right) =\mathbf{X}^{'}$, by the continuity of $c(\cdot)$, we have:
\begin{equation}
\underset{k\rightarrow \infty}{\lim}\left| \mathrm{c}\left( \mathbf{X}_k,\mathbf{D}_{k} \right) -\mathrm{c}\left( \mathbf{X}^{'},\mathbf{D}_0 \right) \right|=0.
\end{equation}
To sum up, we show  $\mathbf{X}^{'}$ is also an optimal solution at point $\mathbf{D}_0$, which contradicts with that $\mathbf{X}_0$ is the unique solution at $\mathbf{D}_0$. Thus, the assumption of discontinuity does not hold and $f^*\left(\cdot \right)$ is continuous when the optimal solution is unique. 

We further show the Hessian of $f^{*}\left( \cdot \right)$ exists almost everywhere. As $\mathbf{X}^*$ is the optimal solution for $\mathbf{D}$, $\mathbf{X}^*$ and $\mathbf{D}$ satisfy the following first-order Karush Kuhn Tucker (KKT) optimality conditions for~\eqref{ac.opf.nlp}:
\begin{eqnarray}
&&\nabla \mathcal{L} =0,\lambda _{i}^{*} g_i\left( \mathbf{X}^*,\mathbf{D} \right) =0, \nonumber\\
&&\lambda _{i}^{*} \ge 0, g_i\left( \mathbf{X}^*,\mathbf{D} \right) \le 0, i\in \mathcal{M},
\label{ac.opf.kkt}
\end{eqnarray}
\noindent where $\mathcal{L} =c\left( \mathbf{X}^*, \mathbf{D}\right) +\sum_{i\in \mathcal{M}\,}{\lambda _{i}^{*}g_i\left( \mathbf{X}^*,\mathbf{D} \right)}$, and $\lambda _{i}^{*}$ is the corresponding optimal Lagrangian multiplier. Let $\mathbf{V}=[\mathbf{X}^*;\lambda _{i}^{*},i\in \mathcal{M}]$ and $\mathcal{L} _{\mathbf{X}^*}$ represent the partial derivatives of $\mathcal{L}$ w.r.t to $ \mathbf{X}^*$. Before proceeding, we first define the following two matrices $A$ and $B$:
\begin{equation}
\small
A=\left[ \begin{matrix}
	\nabla _{\mathbf{X}^*}\mathcal{L} _{\mathbf{X}^*}&		\nabla _{\mathbf{X}^*}g_1(\mathbf{X^*,D})&		...&		\nabla _{\mathbf{X}^*}g_i(\mathbf{X^*,D})\\
	-\lambda _1\nabla _{\mathbf{X}^*}^{T}g_1(\mathbf{X^*,D})&		-g_1(\mathbf{X^*,D})&		...&		0\\
	...&		...&		...&		...\\
	-\lambda _i\nabla _{\mathbf{X}^*}^{T}g_i(\mathbf{X^*,D})&		0&		...&		-g_i(\mathbf{X^*,D})\\
\end{matrix} \right]  \nonumber
\end{equation}
and 
\begin{equation}
\small
B=\left[ \begin{array}{c}
	\nabla _{\mathbf{D}}\mathcal{L} _{\mathbf{X}^*}  \\
	-\lambda _1\nabla _{_{\mathbf{D}}}^{T}g_1(\mathbf{X^*,D})\\
	...\\
	-\lambda _i\nabla _{_{\mathbf{D}}}^{T}g_i(\mathbf{X^*,D})\\
\end{array} \right],\nonumber
\end{equation}
\noindent where $\nabla _{\mathbf{X}^*} g_i(\mathbf{X^*,D})$, $\nabla _{\mathbf{D}} g_i(\mathbf{X^*,D})$, 
$\nabla _{\mathbf{X}^*}\mathcal{L} _{\mathbf{X}^*}$ and $\nabla _{\mathbf{D}}\mathcal{L} _{\mathbf{X}^*}$ denote the partial derivatives of $g_i(\mathbf{X^*,D})$ w.r.t. $\mathbf{X}^*$, the partial derivatives of $g_i(\mathbf{X^*,D})$ w.r.t. $\mathbf{D}$, the partial derivatives of $\mathcal{L} _{\mathbf{X}^*}$ w.r.t $\mathbf{X}^*$ and the partial derivatives of $\mathcal{L} _{\mathbf{X}^*}$ w.r.t $\mathbf{D}$, respectively. By leveraging Theorem 2 in~\cite{spingarn1979generic}, we can derive $A$ is invertable except for a set of Lebesgue measure zero. We further have the first-order derivatives of $\mathbf{V}$ w.r.t. $\mathbf{D}$ (say $\frac{d\mathbf{V}}{d\mathbf{D}}$) equals to $-A^{-1}B$ when $A$ is invertable by the sensitivity analysis results~\cite{jittorntrum1984solution}. To this end, $\frac{d\mathbf{V}}{d\mathbf{D}}$ also exists except for a set of Lebesgue measure zero. As the elements in matrices $A$ and $B$ are differentiable w.r.t $\mathbf{D}$, the second-order derivatives of $\mathbf{V}$ w.r.t the $\mathbf{D}$ exist and can be computed as follows. Suppose $\mathbf{D}_i$ denote the $i$-th dimension of $\mathbf{D}$. For every $\mathbf{D}_i$, the second-order derivatives of $\mathbf{V}$ w.r.t $\mathbf{D}_i$ is:
\begin{eqnarray}
\frac{d^2\mathbf{V}}{d{\mathbf{D}_i}^2}=A^{-1}\frac{dA}{d\mathbf{D}_i}A^{-1}B_i-A^{-1}\frac{dB_i}{d\mathbf{D}_i},
\label{eq31}
\end{eqnarray}
\noindent where $B_i$ denotes the $i$-th column of $B$. $\frac{dA}{d\mathbf{D}_i}$ and $\frac{dB_i}{d\mathbf{D}_i}$ represent the partial derivatives of all elements in $A$ w.r.t. $\mathbf{D}_i$ and the partial derivatives of all elements in $B_i$ w.r.t. $\mathbf{D}_i$, respectively. We note that the second-order derivatives of $\mathbf{X^*}$ w.r.t $\mathbf{D}_i$ can be directly obtained from the elements in the first $4\cdot \text{card}(\mathcal{N})$ rows of $\frac{d^2\mathbf{V}}{d{\mathbf{D}_i}^2}$. Thus, we can leverage~\eqref{eq31} to get the second-order derivatives of $\mathbf{X^*}$ w.r.t every $\mathbf{D}_i$ and combine them to get the Hessian of $f^*(\cdot)$.

\end{proof}

\section{Proof of Theorem~\ref{lower.bound.ac.thm}} \label{thm5_proof}
\begin{proof}
(Sketch) We adopt similar ideas from~\cite{SafranS17} and extend the approximation error for the uniform distribution to general distribution function that is lower-bounded with a positive constant. Specifically, we first derive the one-dimensional lower bound for the expected square error by using piece-wise linear functions for approximating quadratic functions on the interval $[0,1]$ with a positive constant of the density function. We then extend the lower bound to the general nonlinear function with lower-bounded second-order derivatives on a non-zero measure. For the multi-dimensional case, by restricting the target function to a line in a particular direction is non-linear (with Hessian whose smallest eigenvalue is lower bounded by a positive constant), we can integrate along all lines in the direction and use the one-dimensional result to establish the lower bound accordingly.
\end{proof}


\end{appendices}

\end{NoHyper}
\end{document}